\newcommand{\bld}[1]{\mbox{\boldmath$#1$\unboldmath}}
\newcommand{\bov}{{\bld{v}}}
\newcommand{\bk}{{\bld{k}}}
\newcommand{\bkp}{{\bld{k'}}}
\newcommand{\bp}{{\bld{p}}}
\newcommand{\bq}{{\bld{q}}}
\newcommand{\bB}{{\bld{B}}}
\newcommand{\bxp}{{\bld{x_\perp}}}
\newcommand{\bdxp}{{\bld{\Delta x_\perp}}}
\newcommand{\bcdot}{{\bld{\cdot}}}
\newcommand{\bnabla}{{\bld{\nabla}}}
\newcommand{\bwup}{{\bld{w^\uparrow}}}
\newcommand{\bwdown}{{\bld{w^\downarrow}}}
\newcommand{\wup}{{{w^\uparrow}}}
\newcommand{\wdown}{{{w^\downarrow}}}

\newcommand{\bnablap}{{\bld{\nabla_\perp}}}
\newcommand{\bwuppp}{{\bld{w^\uparrow_\perp}}}
\newcommand{\bwdownpp}{{\bld{w^\downarrow_\perp}}}
\newcommand{\wupp}{{{w^\uparrow\!\!}}}

\newcommand{\bwupp}{{\bld{w^\uparrow\!\!}}}
\newcommand{\bwdownp}{{\bld{w^\downarrow}}}
\newcommand{\bwupphk}{{\bld{{w}^\uparrow_{\bld k}\!\!}}}
\newcommand{\bwdownphk}{{\bld{{w}^\downarrow_{\bld k}}}}
\newcommand{\bwupphkp}{{\bld{{w}^\uparrow_{\bld k'}\!}}}
\newcommand{\bwdownphkp}{{\bld{{w}^\downarrow_{\bld k'}}}}
\newcommand{\bwupphmk}{{\bld{{w}^\uparrow_{-\bld k}}}}
\newcommand{\psiupk}{{{{\psi}^\uparrow_{\bld k}}}}
\newcommand{\psidownk}{{{{\psi}^\downarrow_{\bld k}}}}
\newcommand{\psiupp}{{{{\psi}^\uparrow_{\bld p}}}}
\newcommand{\psiupmk}{{{{\psi}^\uparrow_{-\bld k}}}}
\newcommand{\psidownmk}{{{{\psi}^\downarrow_{-\bld k}}}}
\newcommand{\psiupkp}{{{{\psi}^\uparrow_{\bld k'}}}}
\newcommand{\psiupq}{{{{\psi}^\uparrow_{\bld q}}}}

\newcommand{\wupz}{{{w^\uparrow_z}}}
\newcommand{\wdownz}{{{w^\downarrow_z}}}
\newcommand{\zhat}{{{\bld{\hat{z}}}}}
\newcommand{\khat}{{{\bld{\hat{k}}}}}
\newcommand{\pt}{{\partial_t}}
\newcommand{\pz}{{\partial_z}}

\newcommand{\arms}{{A_{\rm rms}}}
\newcommand{\taucor}{{\tau_{\rm corr}}}
\newcommand{\taucas}{{\tau_{\rm cas}}}
\newcommand{\eupp}{{{e^\uparrow}}}
\newcommand{\edownp}{{{e^\downarrow}}}
\newcommand\lp{{\lambda}}

\newcommand\lpar{{\Lambda}}

\newcommand{\wuppl}{{{w^\uparrow_{\lp}}}}
\newcommand{\wdownpl}{{{w^\downarrow_\lp}}}

\newcommand\alphaup{{\alpha}}
\newcommand\alphaupup{{\alpha_\uparrow}}
\newcommand\alphadown{{\alpha_\downarrow}}
\newcommand\kup{{k_2}}
\newcommand\kdown{{k_1}}
\newcommand\tcasup{{t_{\rm cas}^\uparrow}}
\newcommand\tcasdown{{t_{\rm cas}^\downarrow}}
\newcommand\tdiss{{t_{\rm diss}}}
\newcommand\ldiss{{\lambda_{\rm diss}}}
\newcommand\lout{{\lambda_{\rm out}}}
\newcommand{\wupplout}{{{w^\uparrow_{\lout}}}}
\newcommand{\wdownplout}{{{w^\downarrow_\lout}}}
\newcommand\wdiss{{w_{\rm diss}}}
\newcommand\tcas{{t_{\rm cas}}}
\newcommand\epsupt{{\tilde{\epsilon}^\uparrow}}
\newcommand\epsdownt{{\tilde{\epsilon}^\downarrow}}
\newcommand\kout{{k_{\rm out}}}
\newcommand\kmax{{k_{\rm max}}}
\newcommand\kdiss{{k_{\rm diss}}}

\newcommand\lpel{{{l}}}
\newcommand{\wdownplel}{{{w^\downarrow_{l}}}}
\newcommand{\wupplel}{{{w^\uparrow_{l}}}}
\newcommand\zup{{{z^\uparrow}}}

\documentclass[11pt,preprint]{aastex}

\begin{document}
\title{Imbalanced Weak MHD Turbulence}
\author{Yoram Lithwick and Peter Goldreich}
\affil{130-33 Caltech, Pasadena, CA 91125; yoram@tapir.caltech.edu,
pmg@gps.caltech.edu}

\begin{abstract}

MHD turbulence consists of waves that propagate along magnetic
fieldlines, in both directions.  When two oppositely directed waves collide,
they distort each other, without changing their 
respective energies.
In weak MHD turbulence, a given wave suffers many collisions
before cascading.  
``Imbalance'' means that more energy
is going in one direction than the other.
In general, MHD turbulence is imbalanced.  
A number of complications arise 
for the imbalanced cascade that are unimportant
for the balanced one.

We solve weak MHD turbulence that is imbalanced.  
Of crucial importance is that 
the energies going in both
directions are forced to equalize at the dissipation
scale.  We call this the
``pinning'' of the energy spectra.  It affects the entire
inertial range.

Weak MHD turbulence
is particularly interesting because 
perturbation theory is applicable.
Hence it can be described with a simple kinetic equation.
Galtier et al. (2000) derived this kinetic equation. 
We present a simpler, more physical derivation, based 
on the picture of colliding wavepackets.
In the process, we clarify the role of the 
zero-frequency mode. We also explain why Goldreich \& Sridhar 
claimed that perturbation
theory is inapplicable, and why this claim is wrong.  
(Our ``weak'' is equivalent to Goldreich \& Sridhar's ``intermediate.'')

We perform numerical simulations of the kinetic equation to verify
our claims.  
We construct simplified model equations that illustrate
the main effects. Finally, we show that a large magnetic Prandtl
number does not have a significant effect, and that
hyperviscosity leads to a pronounced bottleneck effect.

\end{abstract}

\section {Introduction}

MHD turbulence is ubiquitous in astrophysics.  
For example, it is present in the
sun, the solar wind, the interstellar medium, molecular clouds,
accretion disks, and galaxy clusters.
Theoretical understanding of 
incompressible MHD turbulence has grown
explosively in the last decade.
Nonetheless, it remains underdeveloped.

Iroshnikov (1963) and Kraichnan (1965) developed a
theory for MHD turbulence.  They realized
that the magnetic field at the largest lengthscale in a cascade
directly affects all of the
smaller lengthscales.
Small-scale fluctuations can be treated as 
small-amplitude waves in the presence of a large mean 
magnetic field.  By contrast, the large-scale velocity is
unimportant for small-scale dynamics; it
can be eliminated by a change of variables, since the equations
of MHD are invariant under Galilean transformations.\footnote{
One outgrowth of Iroshnikov and Kraichnan's theory that is particularly
relevant to the present paper is Grappin, Pouquet, and L\'eorat (1983).
We discuss it in \S \ref{sec:sses}.}

Despite their realization 
of the importance of the mean magnetic
field, Iroshnikov and Kraichnan assumed that 
small-scale fluctuations are isotropic.
Numerical simulations later showed that this assumption 
is wrong.  
Even with isotropic excitation at large scales,
fluctuations on smaller scales are elongated along the
mean magnetic field
(e.g., Montgomery \& Turner 1981, 
Shebalin, Matthaeus, \& Montgomery 1983).

In retrospect, it is not very surprising to find elongated
fluctuations.
Arbitrary disturbances in incompressible MHD can be decomposed into 
shear-Alfv\'en and pseudo-Alfv\'en waves.  
Each wave travels either
up or down the mean field at the Alfv\'en speed, $v_A$, which
is the magnitude of the mean field in velocity units.
Consider stirring a magneto-fluid with a spoon that is moving
at speed $v\ll v_A$, 
for a time comparable to the spoon's width divided by
$v$.\footnote{ 
In a turbulent cascade,
one would expect $v\ll v_A$ on small scales, since $v$ decreases towards 
small scales, whereas $v_A$ is unchanged.}
Alfv\'en waves are radiated away from the spoon, 
parallel
to the mean field with speed
$\pm v_A$.
After the disturbance is finished, there are two wavepackets travelling
away from each other.
Each wavepacket is elongated along the mean field, with 
parallel-to-transverse
aspect ratio
$\sim v_A/v\gg 1$.

The characteristics of MHD turbulence depend critically
on the amount of elongation. 
When parallel-to-transverse aspect ratios are smaller than
$v_A/v$, waves collide many times before
cascading.  Hence the turbulence is ``weak,'' and
perturbation theory can be used to derive
a kinetic equation and a spectrum
(Sridhar \& Goldreich 1994; see Zakharov, L'vov, \& Falkovich 1992 
for a general review of
weak turbulence).
Goldreich \& Sridhar (1997) and Ng \& Bhattacharjee (1997) deduced
the spectrum of the balanced 
weak cascade from scaling arguments.\footnote{
We relegate some of the history to a footnote because
it can be confusing.
Sridhar \& Goldreich (1994) developed the first theory
of MHD turbulence that accounted for the anisotropy
of fluctuations.
They claimed that 
``three-wave'' processes vanish in weak MHD turbulence, 
and ``four-wave'' processes
must be considered (i.e., perturbation theory
is trivial to first order, so second order terms are important).
As a result, they used four-wave couplings
to derive a kinetic equation and a spectrum for weak MHD turbulence.
Montgomery \& Matthaeus (1995) claimed,
and Ng \& Bhattacharjee (1996) showed, that Sridhar \& Goldreich (1994) 
are wrong, and three-wave processes do 
not vanish.  Goldreich \& Sridhar (1997) explained
the contradiction:  Sridhar \& Goldreich (1994)
had unknowingly assumed that fieldline wander is limited,
i.e., that the separation between any two fieldlines is nearly 
constant along their entire length;
in this case,
three-wave couplings are negligible and the kinetic equation based
on four-wave couplings is correct.
In the more realistic case that fieldlines do wander, 
three-wave processes are
important.  Goldreich \& Sridhar (1997) went on to argue that,
in the latter case,
perturbation theory is inappropriate, and
couplings of all order are of comparable magnitude; so they called this ``intermediate''
turbulence.
Galtier et al. (2000) argued that perturbation theory is appropriate, even
when three-wave processes are important.  
In the Appendix of the present paper, we use Goldreich \& Sridhar's picture
of wavepackets following wandering fieldlines to clarify the controversy,
and to explain why perturbation
theory works.  Because it does work, we call the cascade ``weak'' instead of 
``intermediate.''}
However, similar scaling arguments are inadequate for the
imbalanced cascade (see \S \ref{sec:insufficiency} of the
present paper).
Galtier et al. (2000) derived
the kinetic equation for the weak imbalanced cascade.
Their balanced spectrum
agrees with that of Goldreich \& Sridhar (1997) and 
Ng \& Bhattacharjee (1997).
They also presented a partial solution for the general
imbalanced case.
In \S \ref{sec:insufficiency},
we explain why their solution is incomplete; in \S \ref{sec:sses}
we give the complete solution.

Even if aspect ratios are smaller than $v_A/v$ on
large scales, at a small enough scale they become comparable to $v_A/v$.
Below this scale perturbation theory breaks down, and weak turbulence
becomes ``strong.''  
Goldreich \& Sridhar (1995) worked out the scalings 
for the balanced strong cascade.
They argued that aspect ratios are comparable to $v_A/v$ at all scales
in the strong regime.
Strong turbulence is difficult, largely because it
is non-perturbative.
Although strong and weak turbulence differ in 
a number of ways, they also share many similarities.  
One of our motivations for studying weak turbulence is to
gain insight into strong turbulence.  
In particular, turbulence in the solar wind 
is observed to be imbalanced; it cannot be understood
without a theory for imbalanced strong MHD turbulence.
Yet this theory is unknown.
In a future paper, 
we will work it out by 
extending the results of the present paper.

\section {Basic Equations} \label{sec:eom}

Ideal incompressible MHD\footnote{
In this paper, we consider only
incompressible MHD turbulence; compressibility 
does not
alter the dynamics very much (Lithwick \& Goldreich 2001).}
is described by the following equations of motion: 
\begin{eqnarray}
\pt \bov+\bov\bcdot\bnabla\bov&=&-\bnabla P+\bB\cdot\bnabla\bB \ , \label{eq:eomv} \\
\pt\bB+\bov\bcdot\bnabla\bB&=&\bB\cdot\bnabla\bov \ , \label{eq:eomb} \\
\bnabla\cdot\bov&=& \bnabla\cdot\bB=0 \ . 
\end{eqnarray}
The density is set to unity; the fluid velocity is $\bov$; the magnetic
field in velocity units is $\bB\equiv$(magnetic field)$/(4\pi)^{1/2}$; the
total pressure is $P\equiv p+B^2/2$, which is the sum of the thermal and magnetic
pressures.  Viscous and resistive terms are neglected in the above equations;
they are important on small scales, and will be included where required.

We decompose the magnetic field into its mean,
$v_A \zhat$, where
$v_A$ is the Alfv\'en speed and $\zhat$ is a unit vector, and
into its fluctuating part
$\bld{b}\equiv \bB-v_A \zhat$.
With this decomposition, the equations of motion may be
written in terms of the Elsasser variables, $\bwup\equiv \bov-\bld{b}$ and
$\bwdown\equiv \bov+\bld{b}$, as follows:
\begin{eqnarray}
\pt \bwup + v_A \pz\bwup=-\bwdown\bcdot\bnabla\bwup-\bnabla P \ , \label{eq:eom1} \\
\pt \bwdown - v_A \pz\bwdown=-\bwup\bcdot\bnabla\bwdown-\bnabla P \ , \label{eq:eom2} \\
\bnabla\bcdot\bwup=\bnabla\bcdot\bwdown=0 \label{eq:eom3} \ . 
\end{eqnarray}
Note that $P$ is not an independent degree of freedom.  Taking the divergence of either
equation (\ref{eq:eom1}) or (\ref{eq:eom2}) yields 
\begin{equation}
P=-\nabla^{-2}(\bnabla\bwup\bld{:}\bnabla\bwdown) \ , \label{eq:p}
\end{equation}
where $\nabla^{-2}$ is the
inverse Laplacian.  
When $\bwdown=0$, $\bwup$ 
propagates undistorted upwards along the mean magnetic field with speed $v_A$.
Similarly, when $\bwup=0$, $\bwdown$ propagates downwards at $v_A$.  
Nonlinear interactions 
occur only between oppositely directed wavepackets.
It is these interactions that are responsible for turbulence.

There are three conserved quantities in incompressible MHD.  Two of these
are immediately apparent from equations (\ref{eq:eom1})-(\ref{eq:eom3}): the
energies of the upgoing and of the downgoing waves, i.e., $(\wup)^2$ and $(\wdown)^2$.
Technically,
these are twice the energy per unit mass.  We refer to them as simply
energies throughout the paper.
These energies are directly related to the total (kinetic
plus magnetic) energy $\propto (\wup)^2+(\wdown)^2$ and to the cross-helicity 
$\propto (\wup)^2-(\wdown)^2$.  
The focus of this paper is turbulence
where the energies in the up and down waves differ, or, equivalently,
where the cross-helicity is non-zero.
The third conserved quantity is magnetic helicity;  however, 
we consider only non-helical turbulence in this paper, 
so helicity does not play a role.

In MHD turbulence, on lengthscales much smaller than the outer
scale, there is 
effectively a strong mean magnetic field that is
due to fluctuations on the largest lengthscales.
Gradients transverse to this mean field 
are much larger than gradients
along it (e.g., 
Shebalin, Matthaeus, \& Montgomery 1983,
Goldreich \& Sridhar 1995, 1997,
Ng \& Bhattacharjee 1996).  
This allows the MHD equations to be slightly simplified.
Denoting transverse components with the symbol
$\bld{\perp}\equiv (x,y)$,
the transverse components of equation
(\ref{eq:eom1}) are
\begin{equation}
\pt \bwuppp + v_A \pz\bwuppp\approx -\bwdownpp\bcdot\bnablap\bwuppp-\bnablap P \ , 
\label{eq:eomrmhd1}
\end{equation}
assuming that $\wdownz \partial_z\bwupp$ is much smaller than
$\bwdownpp\bcdot\bnablap\bwupp$, 
and
equation (\ref{eq:eom3}) is
\begin{equation}
\bnablap\bcdot\bwuppp\approx 0 \ . \label{eq:eomrmhd2}
\end{equation}
We assume that 
the parallel components of $\bwup$ and $\bwdown$ are either
comparable to, or less than,
their respective perpendicular components.
We will see below that
this is typically the case in the inertial range of a turbulent
cascade.
Similarly,
\begin{equation}
\pt \bwdownpp - v_A \pz\bwdownpp\approx -\bwuppp\bcdot\bnablap\bwdownpp-\bnablap P \ , 
\label{eq:eomrmhd3}
\end{equation}

\begin{equation}
\bnablap\bcdot\bwdownpp\approx 0 \ . \label{eq:eomrmhd4}
\end{equation}
If we change $\approx$ to $=$,
equations (\ref{eq:eomrmhd1})-(\ref{eq:eomrmhd4}) form a closed set.
They are called the equations of reduced MHD.  They apply also in 
compressible MHD whenever transverse gradients are larger than parallel 
ones (e.g., Biskamp 1993).  
The main goal of this paper is to solve these equations.  
Although the complete equations are not much more complicated,
it simplifies our discussions to neglect parallel gradients 
relative to perpendicular ones
at the outset.
There are two conserved energies in reduced MHD:
$(w^\uparrow_\perp)^2$ and $(w^\downarrow_\perp)^2$.

The parallel components of equations (\ref{eq:eom1})
and (\ref{eq:eom2}) are
\begin{eqnarray}
\pt \wupz + v_A \pz\wupz\approx -\bwdownpp\bcdot\bnablap\wupz \ , \\
\label{eq:parallel1}
\pt \wdownz - v_A \pz\wdownz\approx -\bwuppp\bcdot\bnablap\wdownz \ ,
\end{eqnarray}
after neglecting parallel gradients relative to transverse
ones, and after assuming that
$\vert\wupz\vert$ and $\vert\wdownz\vert$ are not much smaller than 
$\vert\bwuppp\vert$ and $\vert\bwdownpp\vert$. 
Clearly, $(\wupz)^2$ and $(\wdownz)^2$ are conserved quantities.
The transverse equations describing reduced MHD are unaffected by these parallel 
equations because the former are independent of $\wupz$ and $\wdownz$.
Nonetheless, the parallel equations have observable consequences.

It is conventional to decompose the 
normal modes of linearized incompressible
MHD, $\bwupp$ and $\bwdownp$, into 
shear-Alfv'en and pseudo-Alfv\'en waves. These correspond to
the Alfv\'en and slow waves of compressible MHD.
When perpendicular gradients are much larger than parallel 
ones, $\bwuppp$ and $\bwdownpp$ are nearly equivalent to
shear-Alfv\'en waves; $\wupz$ and $\wdownz$ are nearly
equivalent to pseudo-Alfv\'en waves.

Also observationally relevant is the evolution of a passive scalar $s$,
which satisfies
$\pt s +\bov\bcdot\bnabla s=0$.
In terms of Elsasser variables, and after neglecting
parallel gradients, the passive scalar satisfies
\begin{equation}
\pt s \approx -(1/2)(\bwuppp+\bwdownpp)\bcdot\bnablap s \ .
\label{eq:scalar}
\end{equation}

To avoid a proliferation of subscripts,
in the remainder of this paper we drop the 
$\perp$ from $\bwuppp$ and $\bwdownpp$.  To denote
the parallel components, we use $\wupz$ and $\wdownz$.

\section{Weak MHD Turbulence: Heuristic Discussion}
\label{sec:heuristic}

One of the virtues of weak MHD turbulence is that it can
be analyzed in a mathematically rigorous way with perturbation
theory; this yields a kinetic equation.  
Nevertheless, we begin with a qualitative description, which captures
most of the features of the turbulent cascade. 

\subsection{Scaling Relation}
\label{sec:scaling}

MHD turbulence can be understood from the dynamics of
$\bwupp$ and $\bwdownp$ (eqs. [\ref{eq:eomrmhd1}-\ref{eq:eomrmhd4}]
for reduced MHD, dropping $\perp$ subscripts).
To linear order,  $\bwupp$ is a wave that propagates
up the mean magnetic fieldlines at the Alfv\'en speed, $v_A$;
$\bwdownp$ propagates down at $v_A$.
Each wave perturbs the mean magnetic fieldlines.  
Nonlinear terms describe the interaction
between oppositely directed waves: each wave nearly
follows the fieldlines perturbed by its collision partner.
\footnote{The equation for a scalar quantity $f$ that travels upwards
at speed $v_A$, while following the magnetic fieldlines of the
down-going $\bwdownp$ is $(\pt+v_A\pz+\bwdownp\bcdot\bnablap)f=0$.
Equation (\ref{eq:eomrmhd1}) for the vector $\bwupp$ differs from this because
of the pressure term,
which is required to keep
$\bwupp$ incompressible, while conserving the
energy $(\wupp\ )^2$.  Thus $\bwupp$ does not exactly 
follow the fieldlines of $\bwdownp$.  Nonetheless,  
this deviation does not greatly affect the
behaviour of the turbulence.  Dissipation  
is a second effect that prevents the following of fieldlines. 
In the present discussion, we consider lengthscales that are
sufficiently large
that dissipation can be neglected.}

Consider an upgoing wavepacket 
that encounters a train of downgoing wavepackets. 
As the upgoing wave travels up the length of the downgoing
train, it is gradually distorted.  It tries to follow the
perturbed fieldlines in the downgoing train, but these fieldlines
``wander,'' i.e., the transverse
separation between any two fieldlines changes.  
When the up-wave has travelled through a sufficiently large number
of downgoing wavepackets
that the amount of fieldline wander is 
comparable to the up-wave's transverse size, then the up-wave cascades.

To be quantitative, let each downgoing wave
in the train have 
a typical amplitude $\wdownpl$, a transverse size
$\lp$, and a parallel size $\lpar$, where ``transverse'' and
``parallel'' refer to the orientation relative to the
mean magnetic field.
The most important collisions are between wavepackets
of comparable transverse size (see \S \ref{sec:locality}).
So let the upgoing wave have transverse size $\lp$ as well.

Since each downgoing wavepacket 
has a typical perturbed magnetic field of magnitude
$\sim\wdownpl$ (neglecting the factor of $1/2$), it bends the
fieldlines by the angle $\wdownpl/v_A$; the transverse
displacement of a fieldline through 
this wavepacket is 
$(\wdownpl/v_A)\lpar$; and the wander of two typical
fieldlines through the wavepacket, if
they are initially separated by $\lp$, 
is also $(\wdownpl/v_A)\lpar$.

In weak turbulence,
the wander through a single wavepacket is smaller than
the wavepacket's transverse size, 
\begin{equation}
{\wdownpl\over v_A}\lpar\ll \lp \ \ {\rm and} \ \ \ 
{\wuppl\over v_A}\lpar\ll \lp \ .
\label{eq:inter}
\end{equation}
When these inequalities are not satisfied,
strong turbulence is applicable;  see \S \ref{sec:strong}.
Thus in weak turbulence
an upgoing wavepacket must travel through many downgoing ones
before cascading.
After $N$ downgoing wavepackets,
fieldlines have wandered a distance
$\sim N^{1/2}(\wdownpl/v_A)\lpar$, assuming that 
wavepackets are statistically independent.
The upgoing wavepacket is fully distorted---and hence cascaded---when 
the fieldlines
it is following wander a distance $\lp$, i.e., when
$N\sim(\lp v_A/\lpar\wdownpl)^2$. Since each downgoing wavepacket
is crossed in the time $\lpar/v_A$, the cascade time of
the upgoing wavepacket is
\begin{equation}
\tcasup\sim\Big({\lp v_A\over \lpar\wdownpl}\Big)^2{\lpar\over v_A}\sim
\Big({\lp \over\wdownpl}\Big)^2
{v_A\over\lpar} \ .
\label{eq:cascadetime}
\end{equation}
In this time,
the upgoing wavepacket travels a distance 
$v_A\tcasup$, which is much larger than its own
length, $\lpar$.\footnote 
{We assume throughout this paper that the upgoing waves' parallel
lengthscale is the same as that of the downgoing waves,
$\lpar$; the extension
to the case when they differ is trivial, as long as the inequalities
(\ref{eq:inter}) are both satisfied, with the appropriate $\lpar$'s.}
The head and the tail of the upgoing wavepacket are both distorted
by the same downgoing wavepackets; so both head and tail undergo nearly
the same distortion as they cascade.\footnote{The head
of the upgoing wavepacket slightly distorts each downgoing wavepacket;
so the downgoing wavepacket seen by the tail is slightly distorted relative
to that seen by the head.
Nonetheless, 
this backreaction is a higher-order correction that
can be ignored in weak turbulence;
see \S \ref{sec:prelim}.
}
Consequently, as the upgoing wavepacket cascades to smaller transverse
lengthscales, it does not cascade to smaller parallel ones:
\begin{equation}
\lpar={\rm scale \ independent} \ . \label{eq:intscal1}
\end{equation}
A proof of this follows from the 3-wave resonance relations
(Shebalin, Matthaeus, \& Montgomery 1983; see also
\S \ref{sec:compare} of the present paper).

We calculate the steady state energy spectra by using Kolmogorov's
picture of energy flowing from large to small lengthscales.
The energy in up-waves flows 
from lengthscales larger than
$\lp$ to those smaller than $\lp$ at the rate
\begin{equation}
\epsilon^\uparrow\sim\frac{(\wuppl)^2}{ \tcasup} 
\sim\Big[\frac{\wuppl\wdownpl}{\lp}\Big]^2\frac{\lpar}
{v_A} \label{eq:epsilonheuristic}
 \ . 
\end{equation}
We call this simply the ``flux.''
We define 
$\epsilon^\uparrow$ more precisely below
(eq. [\ref{eq:flux}]). 
In steady state, the flux must
be independent of $\lp$, so
\begin{equation}
\wuppl\wdownpl\propto \lp \ . \label{eq:intscal2}
\end{equation}

\subsection{Insufficiency of Scaling Arguments for the Imbalanced Cascade}
\label{sec:insufficiency}

A balanced cascade has $\wuppl=\wdownpl\equiv w_\lambda$.  
Its solution in steady state
is simple:
$w_\lambda\propto \lp^{1/2}$ and
$\epsilon^\uparrow=\epsilon^\downarrow\sim w_\lambda^4\lpar/\lp^2 v_A$
(Goldreich \& Sridhar 1997, Ng \& Bhattacharjee 1997).
However, if the cascade is imbalanced, 
a number of complications arise.

By the symmetry between up- and down-going waves, 
the down-going flux is given by
the analogue of equation (\ref{eq:epsilonheuristic}): 
\begin{equation}
\epsilon^\downarrow\sim \Big[\frac{\wuppl\wdownpl}{\lp}\Big]^2\frac{\lpar}
{v_A} \ .
\label{eq:epsilondownheuristic}
\end{equation}
Because $\epsilon^\uparrow$ and $\epsilon^\downarrow$ both
depend on the same combination of $\wuppl$ and $\wdownpl$---namely
their product---the steady state solution is non-trivial.
Had this degeneracy not occurred, e.g., had
we found
\begin{equation}
\epsilon^\uparrow\sim 
{(\wuppl)^{2+\gamma}(\wdownpl)^{2-\gamma}\over\lp^2}
{\lpar\over v_A} \ \ {\rm and} \ \ 
\epsilon^\downarrow\sim
{(\wuppl)^{2-\gamma}(\wdownpl)^{2+\gamma}\over\lp^2}
{\lpar\over v_A}  
 \ , 
\end{equation}
where $\gamma\ne 0$, then the solution would have been simple:
$\wuppl\propto\wdownpl\propto\lp^{1/2}$ and $(\epsilon^\uparrow/
\epsilon^\downarrow)\sim(\wuppl/\wdownpl)^{2\gamma}$, which
follow from the constancy of $\epsilon^\uparrow$ and $\epsilon^\downarrow$
with $\lp$.  

But in weak MHD turbulence $\gamma=0$ (eqs. [\ref{eq:epsilonheuristic}] 
and [\ref{eq:epsilondownheuristic}]);
constancy of $\epsilon^\downarrow$ 
with $\lp$
is forced by the constancy of $\epsilon^\uparrow$,
and does not yield new information.
One implication is that 
scaling arguments are insufficient to determine
the flux ratio $\epsilon^\uparrow/\epsilon^\downarrow$.
Physically, any flux ratio should be possible. 
But without the dimensionless coefficients of
equations (\ref{eq:epsilonheuristic}) and
(\ref{eq:epsilondownheuristic}), 
$\epsilon^\uparrow/\epsilon^\downarrow$
cannot be determined.  The coefficients
cannot be obtained from scaling arguments; they
depend on the spectral slopes of $\wuppl$ 
and $\wdownpl$
(which are related through $\wuppl\wdownpl\propto\lp$,
eq. [\ref{eq:intscal2}]).
Galtier et al. (2000) calculated the coefficients using
kinetic equations. We explain how in \S \ref{sec:steadystatefluxes}.  
Therefore these authors were able to
relate the flux ratio to the spectral slopes.

The arguments presented thus far 
are still insufficiently 
constraining.
Equations (\ref{eq:epsilonheuristic}) and
(\ref{eq:epsilondownheuristic}) constrain
only the
product $\wuppl\wdownpl$.
There are seemingly an infinite number of solutions
with given values of
$\epsilon^\uparrow$ and $\epsilon^\downarrow$,
since $\wuppl$ can be multiplied by any constant
as long as $\wdownpl$ is divided by this same constant.
Furthermore, we expect on physical grounds that if the values
of $\wuppl$ and $\wdownpl$ at a given lengthscale 
are fixed (instead of the values of $\epsilon^\uparrow$ and
$\epsilon^\downarrow$), the cascade should 
be completely constrained; however, in this case
the constancy of equations  (\ref{eq:epsilonheuristic}) and
(\ref{eq:epsilondownheuristic}) leaves the $\lambda$-dependence
of $\wuppl/\wdownpl$ completely undetermined---even given
the coefficients derived by Galtier et al. (2000).
Do $\wuppl$ and $\wdownpl$ cross? Are they cut off by dissipation
at the same scale?  
All of these problems for the imbalanced
cascade can be resolved once the dynamics
at the dissipation scale is understood.

\subsection{Dynamics at the Dissipation Scale: Pinned Spectra}
\label{sec:pinned}

The main result of the present paper is that 
the energies of the up- and down-going waves
are forced to equalize---they are ``pinned''---at the 
dissipation scale.  
This completely constrains the cascade.
It is unusual that the dynamics
at the dissipation scale has such an important
influence.
In this subsection 
we explain why pinning occurs. 
In \S \ref{sec:sses}, we give the resulting solution 
of the steady state cascade.

From equation (\ref{eq:cascadetime}), the cascade
time of the up-going waves
is inversely proportional to the energy of
the down-going ones:
$\tcasup=
(\lp/\wdownpl)^2
(v_A/\lpar)$, and similarly for the downgoing waves.
We consider how the spectra evolve if initially, on lengthscales
comparable to the dissipation scale,  waves going
in one direction are more energetic than the oppositely
directed ones.  To facilitate the discussion, we refer
to Figure \ref{fig:ke1} on page \pageref{fig:ke1}, 
which presents the results from
a numerical simulation that we discuss in detail in
\S \ref{sec:fixedenergy}.
In the middle panels of Figures \ref{fig:ke1}a-d, we plot
$\eupp(k)\sim(\lp\wuppl)^2$ and $\edownp(k)\sim(\lp\wdownpl)^2$
as functions of wavenumber $k=1/\lp$.
The initial condition is shown in Figure
\ref{fig:ke1}a.
Initially,
$\wuppl>\wdownpl$,
so $\tcasup>\tcasdown$.  We
consider a lengthscale-dependent dissipation time,
$\tdiss$, that is the same for both up- and down-going
waves.\footnote{For example, in the simulation presented
in Figure \ref{fig:ke1},
$\tdiss\simeq\lp^2/\nu$, where $\nu$ is both the viscosity and
the resistivity.} On large lengthscales, the
dissipation timescale is much longer than the cascade times;
$\tdiss$ decreases faster
with increasing $k$ than both $\tcasup$ and $\tcasdown$.
The effects of dissipation are felt on lengthscales
where $\tdiss$ is comparable to---or smaller than---either
$\tcasup$ or $\tcasdown$.
Since $\tcasup>\tcasdown$ in the vicinity
of the dissipation scale, the largest
lengthscale at which dissipation effects are felt is where
$\tcasup\sim\tdiss$.  In Figure \ref{fig:ke1}a, this is
at $k\sim 4,000$.
We now let the spectra evolve; see Figs. \ref{fig:ke1}b-c.
We hold the energies at $k\simeq 1$ fixed; this does not
affect the short-time behaviour shown in Figs. \ref{fig:ke1}b-c.
Since $\wuppl$ feels the dissipative effects at $k\sim 4,000$,
its spectrum is exponentially cut off at smaller scales.  This implies
that the cascade time of the down waves, $\tcasdown$,
increases exponentially towards smaller scales.
As a result, down-wave
energy that is being cascaded from large to small scales
cannot be cascaded fast enough at $k\gtrsim 4,000$.
Therefore the down-waves' energy flux is backed up, and the
$\wdownpl$ spectrum rises.  Furthermore, as $\wdownpl$ rises, $\tcasup$
falls, so the cascade time of the up-waves on small scales decreases, and the
up-wave spectrum falls.  The final result is that the two spectra
are pinned at the dissipation scale.  This pinning
occurs very quickly: on the dissipation timescale.

\section{Steady State Energy Spectra}
\label{sec:sses}

In steady state, the energy spectra $\wuppl$
and $\wdownpl$ are power laws that 
(i) are pinned at the dissipation scale, and 
(ii) 
satisfy  $\wuppl\wdownpl\propto \lp$ 
(eq. [\ref{eq:intscal2}]).
These two conditions completely characterize the
steady state spectra, 
as long as
$\wuppl$ and $\wdownpl$ are specified at the 
outer scale.
It is a remarkable feature of weak MHD turbulence that the dynamics
at the dissipation scale dramatically affects the entire cascade.
Normally, the energy in turbulent cascades is 
viewed as flowing unimpeded
from larger scales to smaller scales. Energy
is injected at the outer scale and swallowed up at the dissipation scale.
But in weak MHD turbulence, if initially the spectra are not 
pinned, then the spectrum with lower energy becomes backed
up, and its energy increases until pinning occurs.

A similar effect was found by
Grappin, Pouquet, and L\'eorat (1983).\footnote{
We thank Bill Matthaeus for pointing out this reference
to us.}
These authors modelled imbalanced MHD turbulence with
an EDQNM closure approximation.   
However, they assumed that the turbulence is isotropic,
which is incorrect.  Nonetheless, they found in their model that
scaling arguments constrained
only the sum of the slopes of the up- and down-going waves' spectra, and
that the two spectra are constrained to be equal
at the dissipation scale.
Thus they discovered pinning, even though they analyzed an invalid
model of MHD turbulence.

In the remainder of this section, we derive the
scalings in steady state.
We denote the dissipation scale by $\ldiss$, and
the value of $\wuppl$ and $\wdownpl$ at $\ldiss$
by $\wdiss$.
We can then express the energy spectra as follows
\begin{eqnarray}
\wuppl&=&\wdiss\Big(
{\lambda\over\ldiss}
\Big)^{(1+\alphaup)/2} \label{eq:spec1} \\
\wdownpl&=&\wdiss\Big(
{\lambda\over\ldiss}
\Big)^{(1-\alphaup)/2} \ . \label{eq:spec2}
\end{eqnarray}
These spectra are valid in the ``inertial range,'' i.e., on
lengthscales larger than the dissipation scale, and smaller
than the outer scale.
There are three parameters that must be calculated  
to constrain the spectra: $\wdiss$, $\ldiss$, and $\alphaup$.

For definiteness, we assume that
dissipation is caused by a diffusive process, described by a
viscous term of the form $\nu\bnabla^2\bov$ 
in equation (\ref{eq:eomv})
and a resistive term of the form $\nu\bnabla^2\bB$
in equation (\ref{eq:eomb}).
This implies that the magnetic Prandtl number is equal to one.\footnote{
It is straightforward to consider other forms for the
dissipation. We consider magnetic Prandtl numbers greater than
unity in \S \ref{sec:prandtl}, and hyperviscosity in \S \ref{sec:bottle}.}
The dissipative timescale is
\begin{equation}
\tdiss(\lp)\simeq\frac{\lambda^2}{\nu} \ .
\label{eq:tdiss}
\end{equation}
At the dissipation scale, the cascade time
is equal to $\tdiss$, i.e., 
$\tcas(\ldiss)=\tdiss(\ldiss)$,
which implies that
\begin{equation} 
\frac{\ldiss^2}{\nu} 
\simeq 
\Big({\ldiss \over\wdiss}\Big)^2
{v_A\over\lpar} \ ,
\end{equation}
after using equations 
(\ref{eq:cascadetime}) and (\ref{eq:tdiss}); so
\begin{equation}
\wdiss\simeq\Big({\nu v_A\over \lpar}\Big)^{1/2} \ .
\label{eq:wdiss}
\end{equation}
Thus when the dissipation is caused by a diffusive process,
$\wdiss$
is independent of the outer-scale energies and the inertial-range
fluxes.  This is not
true for $\ldiss$ or $\alphaup$.

To calculate $\ldiss$ and $\alphaup$, we consider two alternative
scenarios: specified energies at the outer scale,
and specified fluxes.

\subsection{Fixed Energies at the Outer Scale}
\label{sec:fixedenergies}

Suppose that the energies are specified
at the outer scale, $\lout$, where $\wuppl$ and $\wdownpl$
are denoted by $\wupplout$ and $\wdownplout$. 
Since $\wuppl\wdownpl/\lp=\wdiss^2/\ldiss$,
\begin{equation}
\ldiss\simeq{\lout \over\wupplout\wdownplout}{\nu v_A\over \lpar} \ ,
\label{eq:ldiss}
\end{equation}
after using equation (\ref{eq:wdiss}).

Dividing equation (\ref{eq:spec1}) by equation
(\ref{eq:spec2}) yields
$\wupplout/\wdownplout=(\lout/\ldiss)^\alphaup$, so
\begin{equation}
\alphaup=
{\ln \big[\wupplout/\wdownplout\big] \over
\ln \big[\lout/\ldiss\big]}\simeq 
{\ln \big[\wupplout/\wdownplout\big] \over
\ln \big[\wupplout\wdownplout \lpar/(\nu v_A)\big]} 
\ . 
\label{eq:alphaup}
\end{equation}
With $\wupplout/\wdownplout$ fixed, the cascade is
balanced ($\alphaup\rightarrow 0$) in the limit
that the 
inertial range is infinitely large 
($\lout/\ldiss\rightarrow \infty$).

Inserting equations (\ref{eq:wdiss}), (\ref{eq:ldiss}),
and (\ref{eq:alphaup}) into the spectra 
(eqs. [\ref{eq:spec1}] and [\ref{eq:spec2}]) gives
the solution to the steady state imbalanced weak
cascade, assuming that $\wupplout$ and $\wdownplout$
are specified.

Although we have solved for the spectra,
recall from \S \ref{sec:insufficiency} that
heuristic arguments are insufficient for calculating
the ratio of the fluxes that are carried by these spectra,
$\epsilon^\uparrow/\epsilon^\downarrow$.
In the following subsection, we show how
$\epsilon^\uparrow/\epsilon^\downarrow$ 
is related to the spectra.
This relation is particularly important for solving
the inverse problem: given the fluxes
$\epsilon^\uparrow$ and
$\epsilon^\downarrow$, what are  
the spectra of $\wuppl$ and $\wdownpl$?

\subsection{Fixed Fluxes}
\label{sec:fixedfluxes}

Physically, we expect that when the fluxes
are fixed, the cascade should be completely constrained.
In this subsection, we solve for the spectra
given $\epsilon^\uparrow$ and
$\epsilon^\downarrow$.  
To accomplish this,
the dimensionless
coefficients of  
equations (\ref{eq:epsilonheuristic}) and
(\ref{eq:epsilondownheuristic}) are required.
For a given power-law solution, 
$\wuppl\propto \lp^{(1+\alphaup)/2}$ and 
$\wdownpl\propto \lp^{(1-\alphaup)/2}$, these coefficients 
depend on $\alphaup$:
\begin{eqnarray}
\epsilon^\uparrow&=&f(\alphaup)
\Big[\frac{\wuppl\wdownpl}{\lp}\Big]^2\frac{\lpar}
{v_A}  \ , \label{eq:epsf1} \\ 
\epsilon^\downarrow&=&f(-\alphaup)
\Big[\frac{\wuppl\wdownpl}{\lp}\Big]^2\frac{\lpar}
{v_A}  \ \label{eq:epsf2},
\end{eqnarray}
where $f(\alpha)$ is a dimensionless 
function of $\alpha$ that must
be calculated.  By the symmetry between up and
down waves, $\epsilon^\uparrow$ and 
$\epsilon^\downarrow$ are both proportional to the same
function $f$, evaluated at $\pm\alpha$.
Since heuristic arguments are insufficient to calculate
the function $f$, it is fortunate that weak
turbulence can be analyzed with perturbation theory.  
Galtier et al. (2000) computed $f$.  We compute
it in equation (\ref{eq:fappendix}) in the Appendix.

The ratio of the fluxes is related to $\alphaup$ by
\begin{equation}
{\epsilon^\uparrow\over\epsilon^\downarrow}
={f(\alphaup)\over f(-\alphaup)} \ .
\label{eq:fluxratio}
\end{equation}
The limit $\vert\alphaup\vert\ll 1$ is particularly 
interesting.
For given outer-scale energies,
if the inertial range is very large then 
the
steady state cascade is nearly balanced
(see the discussion below eq. [\ref{eq:alphaup}]), 
and $\vert\alphaup\vert\ll 1$.
In this limit, 
we show in the Appendix
that
$f(\alphaup)\simeq f(0)\cdot (1+0.5\alphaup)$  
(see eq. [\ref{eq:pt5}]).
Thus
\begin{equation}
{\epsilon^\uparrow\over
\epsilon^\downarrow}-1 \simeq \alphaup\ \ , \ \ \vert\alphaup\vert\ll 1 \ .
\label{eq:alphaupapprox}
\end{equation}
To linear order in $\alphaup$,
the product of the 
fluxes is independent of $\alphaup$: 
\begin{equation}
\epsilon^\uparrow\epsilon^\downarrow\simeq
[f(0)]^2\Big[\frac{\wuppl\wdownpl}{\lp}\Big]^4\Big(\frac{\lpar}
{v_A}\Big)^2 
=[f(0)]^2\Big[\frac{\wdiss^2}{\ldiss}\Big]^4\Big(\frac{\lpar}
{v_A}\Big)^2
\ \ , \ \ \vert\alphaup\vert\ll 1 \ .
\label{eq:epsprod}
\end{equation}
In the Appendix, we show that $f(0)=1.87$
(eq. \ref{eq:f0}]).
Although for most purposes the precise value of $f(0)$ is 
unimportant, we shall need it when discussing our numerical 
simulations.

To summarize, in the limit of small $\vert\alphaup\vert$,
if $\epsilon^\uparrow$ and $\epsilon^\downarrow$ are
specified, then 
the spectra of $\wuppl$ and $\wdownpl$ are given by equations
(\ref{eq:spec1}) and (\ref{eq:spec2}), with $\wdiss$, $\alphaup$,
and $\ldiss$  given
by equations (\ref{eq:wdiss}), (\ref{eq:alphaupapprox}), 
and (\ref{eq:epsprod}). 
Note that, to first order in 
$\alphaup$, the only relation that depends
on the kinetic equation is
equation (\ref{eq:alphaupapprox}).\footnote{Aside from the uninteresting
dependence of equation (\ref{eq:epsprod}) on $f(0)$.}

If, instead of specifying $\epsilon^\uparrow$ and $\epsilon^\downarrow$,
we specify the outer scale energies---in which
case the spectra are given in \S \ref{sec:fixedenergies}---then
equations (\ref{eq:alphaupapprox}) and (\ref{eq:epsprod})
give the resulting fluxes.

\subsection{Three Peripheral Issues}

This subsection may be skipped on a first reading, as it does
not impact
our main line of argument.

\subsubsection{Locality}
\label{sec:locality}

In \S \ref{sec:scaling}, it is assumed that 
the dominant interactions are those 
between wavepackets that have
comparable transverse lengthscales,
i.e., interactions
are ``local'' in lengthscale.  
In this section, we justify this assumption.

We focus on the cascading of an upgoing wavepacket
by downgoing ones.
Let the upgoing wavepacket have transverse size $\lp$, and
let the downgoing wavepackets 
each have transverse size $\lpel$, parallel size $\lpar$,
and amplitude $\wdownplel$.
The upgoing wavepacket 
cascades when the fieldlines
it is following wander a distance comparable 
to its transverse size, $\lp$.   

We consider first the case that $\lpel<\lp$.
Two fieldlines that are separated by $\lp$ at the
head of the downgoing wavepackets wander independently
of each other.  Their transverse separation after
$N$ downgoing wavepackets increases by
\begin{equation}
N^{1/2}(\wdownplel/v_A)\lpar \  \ , \ \ \ \lpel<\lp  \ .
\label{eq:wander1}
\end{equation}

Conversely, if $\lpel>\lp$, then 
the magnetic field can be expanded to linear order
in $\lp$, so the
magnetic field
at two points separated by $\lp$ differs
by $\sim\wdownplel\lp/\lpel$.
Consequently, the
fieldlines separate by 
\begin{equation}
N^{1/2}(\wdownplel/v_A)(\lp/\lpel)\lpar \ \ , \ \ \ \lpel>\lp \  ,
\label{eq:wander2}
\end{equation}
so long as this separation is smaller
than $\lp$.

For 
interactions to be local, i.e., for 
the amount of fieldline wander seen by an up-wave of
transverse size $\lp$ 
to be maximized by those down-waves that have $\lpel\sim\lp$,
the following two conditions must hold: 
(i) $\wdownplel$ is an increasing function of
$\lpel$ (eq. [\ref{eq:wander1}]); and (ii) $\wdownplel/\lpel$
is a decreasing function of $\lpel$ (eq. [\ref{eq:wander2}]). 
So the cascade is local if
\begin{equation}
0<{d\ln\wdownpl\over d\ln\lp}<1 \ .
\label{eq:intercond}
\end{equation}
The same condition clearly holds for $\wuppl$.
In terms of the steady-state scalings,
$\wuppl\propto \lp^{(1+\alphaup)/2}$ and 
$\wdownpl\propto \lp^{(1-\alphaup)/2}$,
the condition
\begin{equation}
-1<\alphaup<1 \ 
\label{eq:alphacond}
\end{equation}
is required
for the cascade to be local; otherwise, nonlocal effects
are important.  
Galtier et al. (2000) derived the inequalities in (\ref{eq:alphacond})
from their kinetic equation.

There is a second reason why the condition 
${d\ln\wuppl/d\ln\lp}<1$ is necessary for
our heuristic arguments to be valid. 
Consider a 
single Fourier mode with wavelength $\lpel$ and amplitude $\wupplel$;
then the typical difference in $w^\uparrow$ between two points separated by 
$\lp<\lpel$ is $\sim\wupplel \lp/\lpel$, expanding
to linear order in $\lp$.
So, if the condition ${d\ln\wuppl/d\ln\lp}<1$ is violated, then
the contribution to
$\wuppl$ (i.e., to the 
typical difference
in $w^\uparrow$ between two points separated by $\lp$)
is dominated by those Fourier modes that have wavelengths $\lpel\gg\lp$.
This is
a different kind of non-locality than considered
previously: 
the upgoing energy that crosses lengthscale $\lp$ comes
directly from much larger scales (with $\lpel\gg \lp$).

\subsubsection{Transition to Strong Turbulence}
\label{sec:strong}

Weak turbulence is applicable when $\wdownpl\lpar/v_A\ll \lp$
and $\wuppl\lpar/v_A\ll \lp$ (eq. [\ref{eq:inter}]).  
Since $\lp$ decreases faster than both $\wdownpl$ and $\wuppl$
(eq. [\ref{eq:intercond}]), even if these inequalities 
are satisfied at large lengthscales, they are violated
at small ones.  Thus weak turbulence has a limited
inertial range.  

Strong turbulence, which is applicable when the
above inequalities are violated, is
of greater relevance 
than weak turbulence
for describing astrophysical 
sites such as the solar wind. 
In strong turbulence 
the change in fieldline separation within
a single wavepacket is not smaller than the transverse
size of the wavepacket. 
This has two implications.
First, equation (\ref{eq:cascadetime}) for the cascade
time is no longer valid; and second, the parallel size of wavepackets
$\lpar$ decreases towards smaller scales because the head
and tail of a wavepacket are independently cascaded.
The balanced strong cascade is worked out by Goldreich \& Sridhar (1995);
we discuss the imbalanced strong cascade in a future paper.  
Strong turbulence is more difficult to analyze than weak turbulence
because it does not submit to perturbation theory.  Nonetheless, a number
of the features of weak turbulence that we develop
in the present paper are applicable to
strong turbulence.  This is one of our motivations for studying 
the weak cascade. 

Since this paper is concerned with weak turbulence, we
choose the dissipation scale to be sufficiently large that
the entire cascade is weak, i.e., 
we require that
\begin{equation}
{w_{\lambda_{\rm diss}}^\uparrow \lpar\over v_A \lambda_{\rm diss}}\ll 1 \ ,
\end{equation}
and similarly for $w_{\lambda_{\rm diss}}^\downarrow$.
Using equations (\ref{eq:wdiss}) 
and (\ref{eq:ldiss}), we can re-write this condition as follows:
\begin{equation}
{\wupplout\wdownplout\over v_A^2}
\ll
{\lout\over\lpar} 
\Big( {\nu\over\lpar v_A}\Big)^{1/2}
\ .
\end{equation}
There is also a lower limit on the product of the outer scale energies, set by
the requirement that the dissipation scale be smaller than the
outer scale, which implies that 
\begin{equation}
{\wupplout\wdownplout\over v_A^2}
\gg
{\nu\over\lpar v_A} \ ,
\end{equation}
(see eq. [\ref{eq:ldiss}]). 
Typically, one expects that $\lpar\sim\lout$, so 
as long as $\nu\ll \lpar v_A$,
the above two inequalities
can be satisfied with appropriately chosen $\wupplout\wdownplout$.

\subsubsection{The Steady State Spectra of $w_z^\uparrow$
and $w_z^\downarrow$ and of a Passive Scalar}

In \S \ref{sec:eom} we showed that the parallel 
components $w_z^\uparrow$ and
$w_z^\downarrow$ do not affect the evolution of 
the perpendicular components, 
$\bwupp$ and $\bwdownp$.\footnote{Recall that we
drop the $\perp$ label.} 
By contrast, the perpendicular components control
the evolution of the parallel ones.  

The steady state spectra of the
parallel components can be derived as follows.  
By analogy between the equation for $w_z^\uparrow$ 
(eq. [\ref{eq:parallel1}]) and that for 
$\bwupp$ 
(eq. [\ref{eq:eomrmhd1}]), 
the cascade time of
$w_z^\uparrow$  is similar to that of
$\bwupp$, i.e., it is given by $\tcasup$
(see eq. [\ref{eq:cascadetime}]). 
We denote the typical amplitude of the parallel
component of the up-wave on lengthscale $\lambda$
by $w_{z,\lambda}^\uparrow$.
The cascade rate of the energy of the parallel
component is $(w_{z,\lambda}^\uparrow)^2/\tcasup$;
in steady state it must be independent of $\lambda$.
Comparing this to the cascade
rate of the perpendicular component,
$(w_\lambda^\uparrow)^2/\tcasup$, 
which must also be independent of $\lambda$,
we deduce
\begin{equation}
w_{z,\lambda}^\uparrow\propto w_\lambda^\uparrow \ .
\label{eq:parspec}
\end{equation}
Note that the evolution equation for $w_z^\uparrow$ 
(eq. [\ref{eq:parallel1}]) is linear.  So 
the overall amplitude of the $w_{z,\lambda}^\uparrow$
spectrum is arbitrary; more accurately, it is set by the 
value of $w_{z,\lambda}^\uparrow$ at the outer scale.
Of course, a similar relation holds for the down-waves: 
\begin{equation}
w_{z,\lambda}^\downarrow\propto w_\lambda^\downarrow \ .
\end{equation}

Finally, we consider the spectrum of a passive scalar, $s$,
whose evolution is described by equation (\ref{eq:scalar}).
From this equation, it is apparent that
the cascade time of the passive scalar is given by
either $\tcasup$ or $\tcasdown$, whichever is shorter.
So, by the same reasoning that we used in deriving equation
(\ref{eq:parspec}), we deduce
\begin{equation}
s_\lambda\propto {\rm min}(\wuppl,\wdownpl) \ ,
\end{equation}
where $s_\lambda$ is the typical value of $s$ on lengthscale
$\lambda$.
Because of the pinning of the spectra, the lesser of 
the two spectra $\wuppl$ and $\wdownpl$ is also the flatter.
So $s_\lambda$ is proportional to the flatter of 
$\wuppl$ and $\wdownpl$.

\section{Kinetic Equations in Weak Turbulence}

Weak turbulence can be analyzed with 
perturbation theory. 
As a consequence, the
evolution of the energy spectra of the up- and down-waves
is described by a closed set of two equations;
in other
words, the two-point correlation functions evolve independently
of all higher-order correlation functions.  This is a great
simplification.   

Evolution equations for the energy spectra---the ``kinetic 
equations''---were
obtained in Galtier et al. (2000, 2001).  We
present an alternate, more physical, derivation in 
the Appendix.  
Such a derivation is useful because it clears up a number of
erroneous claims that have been made in the literature---in particular,
the claim of Goldreich \& Sridhar (1997) that perturbation theory
is inapplicable. 
 
In the following, we summarize the result derived in the 
Appendix.
The kinetic equations are given in Fourier-space.  
We Fourier transform $\bwupp(x,y,z,t)$ 
and $\bwdownp(x,y,z,t)$ in $x$ and $y$ (but not $z$), and
denote the transforms by
$\bwupphk(z,t)$ and $\bwdownphk(z,t)$,
where $\bk$ is purely transverse ($k_z\equiv 0$).
We define  the energy spectra $\eupp$ and $\edownp$ such that
\begin{eqnarray}
\left<\bwupphk(z,t)\bcdot\bwupphkp (z,t)\right>&=&\eupp(k,t)\delta(\bk+\bk') \\
\left<\bwdownphk(z,t)\bcdot\bwdownphkp (z,t)\right>&=&\edownp(k,t)\delta(\bk+\bk') \ ,
\end{eqnarray}
where $\delta(\bk)$ is a two-dimensional Dirac delta-function, and
angled brackets denote both an ensemble average and 
an average over
$z$. (We assume that the turbulence is homogeneous in $z$.) 
Both $\eupp$ and $\edownp$ are real.
The turbulence is isotropic in the transverse plane, so
$\eupp$ and $\edownp$ are
functions of the magnitude of $\bk$.

From equation 
(\ref{eq:kineticequationappendix}) in the Appendix, 
the kinetic equation for the
up-waves is
\begin{equation}
\frac\partial{\partial t}\Big\vert_k
\eupp(k,t)=
{\lpar\over v_A}
k^2
\int_0^\infty d k_2  k_2^3
\Big[\eupp(k_2,t)-\eupp(k,t)\Big]
\int_0^{2\pi} d\theta \sin^2\theta\cos^2\theta
\frac {\edownp(k_1,t)}{k_1^2}-\nu k^2\eupp(k,t) \label{eq:kineticequation}
\end{equation}
where
\begin{equation}
k_1\equiv (k^2+k_2^2-2kk_2\cos\theta)^{1/2} \ .
\label{eq:theta}
\end{equation}
We have included a term
describing diffusive dissipation on small lengthscales,
$-\nu k^2\eupp$, where $\nu$ is the viscosity, which
is assumed to be equal to the  
resistivity.

Because of the symmetry between up- and down-waves, 
the equation for $\edownp$ is the same as that for $\eupp$, but 
with $\uparrow$ and $\downarrow$ everywhere switched.
The steady-state relation between the energy and the flux 
that we use above (eq. [\ref{eq:epsf1}]) 
is obtained in the Appendix by setting the right-hand side of 
equation (\ref{eq:kineticequation}) zero.

Within an order-unity factor,
$k^2\eupp\sim (\wuppl)^2$ when $\lp= 1/k$.
Recall from 
\S \ref{sec:heuristic} that $\wuppl$ is
the ``typical'' value of $\bwupp$ on lengthscale $\lp$;
more precisely, it can be defined as the square root of the second-order structure 
function of $\bwupp$.
Therefore the steady-state scaling $\wuppl\propto\lp^{(1+\alphaup)/2}$ 
(eq. [\ref{eq:spec1}]) corresponds to $\eupp\propto k^{-(3+\alphaup)}$;
similarly, $\wdownpl\propto\lp^{(1-\alphaup)/2}$
corresponds to $\edownp\propto k^{-(3-\alphaup)}$.

\subsection{Energy and Flux in the Kinetic Equation}
\label{sec:energyandflux}

The energy per unit mass in up-waves is 
$\left<\vert\bwupp\vert^2\right>/2=
(1/2)(2\pi)^{-4}\int \eupp d^2\bk$.
So $\eupp$ is proportional to the energy per unit $d^2\bk$.
Conservation of up-wave energy implies that  
$(d/dt)\int (k\eupp)dk=0$, assuming isotropy and 
neglecting dissipation.
This can be seen immediately from equation (\ref{eq:kineticequation})
if we re-write it
as follows (without the diffusive term),
\begin{equation}
(\partial/\partial t)(k e^\uparrow_k)=
(\lpar/v_A)\int_0^\infty (e^\uparrow_{k_2}-e^\uparrow_k) S^\downarrow(k,k_2)dk_2 \ ,
\end{equation}
since $S^\downarrow$ is symmetric,
\begin{equation}
S^\downarrow(k,k_2)\equiv (kk_2)^3 \int_0^{2\pi} \sin^2\theta\cos^2\theta
\ e^\downarrow_{k_1} k_1^{-2}=
S^\downarrow(k_2,k) \ ,
\label{eq:s}
\end{equation}
and $e^\uparrow_2-e^\uparrow_k$ is
antisymmetric.
In comparing our results with those of Galtier et al. (2000),
note that these authors use the energy per unit $k$,
which they denote $E^+$, i.e., $E^+\sim k\eupp$, within
a multiplicative constant. 

The energy flux $\epsilon^\uparrow$ 
is the net rate at which energy flows
across a given wavenumber $k$.
We define it as the integral
of the right-hand side of equation (\ref{eq:kineticequation})
over $d^2\bk/(2\pi)$, from some particular $k$ to infinity,
so that the kinetic equation can be written as
\begin{equation}
(\partial/\partial t)(k e^\uparrow_k)=-
(\partial/\partial k)\epsilon^\uparrow \ .
\label{eq:kineticflux}
\end{equation}  
Explicitly,
we can write the flux in the form
\begin{equation}
\epsilon^\uparrow=\int_0^{k}dp\int_{k}^\infty dq
(e^\uparrow_p-e^\uparrow_q) S^\downarrow(p,q) \ .
\label{eq:flux}
\end{equation}
It is trivial to verify that $\partial/\partial k$ of this expression is
equal to minus the right-hand side of equation (\ref{eq:kineticequation}).
The positive term in equation (\ref{eq:flux}) is the rate at which energy
flows from wavenumbers smaller than $k$ to those greater than $k$;
the negative term is similar, but with the origin and destination 
of the energy switched.

\section{Numerical Simulations of Kinetic Equations}
\label{sec:numkin}

In this section we present the results of numerical simulations
which verify our previous heuristic discussions: in particular, 
the pinning of spectra and the scaling of the spectra in steady state.

It is much faster to simulate kinetic equations than the full
equations of motion for $\bwup$ and $\bwdownp$.  There
are a number of reasons for this. First, the kinetic equations
are only one dimensional, since homogeneity has been assumed
parallel to the mean magnetic field, and isotropy has been
assumed in the plane transverse to the mean magnetic field.
Second, and more importantly, the averaged energies $\eupp$ and
$\edownp$ are much smoother functions of $k$ than $\bwup$
and $\bwdownp$.  Thus, a logarithmically-spaced
grid can be used, which greatly reduces the number of variables that
need to be evolved.  The result is an enormous reduction in
computational time.  A typical kinetic simulation
takes a few hours on a PC to reach steady state.
A comparable fully three-dimensional
MHD simulation, 
would require many months, if not years, on the fastest supercomputers.

Galtier et al. (2000) perform numerical simulations
of the kinetic equation.  However, their investigation
of the imbalanced cascade is incomplete.  In particular,
they only plot spectra of the product $\eupp\edownp$.  They
do not discuss the pinning of the spectra at the dissipation
scale, which is crucial to the evolution of the cascade.

For our numerical simulations of the kinetic equations
(eq. [\ref{eq:kineticequation}] and the analogous
$\edownp$ equation),
we set
\begin{equation}
{\lpar\over v_A}={1\over 2}  \ .
\label{eq:lpar}
\end{equation}
This corresponds to
absorbing $2\lpar/v_A$ into $\eupp$ and $\edownp$.
Our method of integration is very similar to that of
Galtier et al. (2000).
We change variables in equation (\ref{eq:kineticequation}) 
to give
\begin{equation}
\frac\partial{\partial t}\Big\vert_k
\eupp(k,t)=
k
\int_0^\infty d k_2  k_2^2
\Big[\eupp(k_2,t)-\eupp(k,t)\Big]
\int_{\vert k-k_2\vert}^{\vert k+k_2\vert} dk_1 \sin\theta\cos^2\theta
\frac {\edownp(k_1,t)}{k_1} -\nu k^2 \eupp(k,t) \ ,
\label{eq:kincode}
\end{equation}
where $\theta$ is a function of $k_1$ and $k_2$
(given in eq. [\ref{eq:theta}]).
We evaluate all functions of $k$ on 
a fixed, logarithmically-spaced grid 
with $k=2^{i/8}$, $i=1,...,100$. At the outer
scale, 
\begin{equation}
\kout=2^{1/8}=1.09 \ ,
\label{eq:outer}
\end{equation}
and the maximum $k$ is
\begin{equation}
\kmax=2^{100/8}=5793 \ . 
\end{equation}
The double integral over $k_1$ and $k_2$
is performed by summing the values of
the integrand evaluated on the $k$-space grid.  The factor
$\sin\theta\cos^2\theta$
is averaged in the vicinity
of each grid point, i.e., at each ($\kdown$,$\kup$).  Since
$\theta$ is a function of $\kdown/k$ and $\kup/k$, and
since the grid is logarithmic, the averaged angular
factor can be precomputed and stored in a two-dimensional
matrix, each element of which is the average
in the vicinity of ($\kdown/k$, $\kup/k$).\footnote{
In evaluating the integral near $\kout$, 
the values of $\eupp(\kup<\kout)$ and $\edownp(\kdown<\kout)$
are required.  Since these values are off
of the grid, some method of extrapolation is needed.
In the runs that we present in this paper, we extrapolate
with $\eupp,\edownp\propto k^{-2}$.   
With this extrapolation, there
is no energy transfer from
modes with $k<\kout$ to those with $k>\kout$; steeper
extrapolations would transfer energy.
We have also
experimented with a flatter spectrum for $k<\kout$:
$\eupp=\eupp(\kout)$, $\edownp=\edownp(\kout)$.
While this changes the behaviour near $k\sim \kout$---in particular,
it leads to a sharp drop in $\eupp$ and $\edownp$ between
$k=\kout$ and $k=2^{1/8}\kout$ by a factor of around
3---the remainder of the spectrum for $k>2^{1/8}\kout$ is
nearly unaffected.  With the $k^{-2}$ extrapolation there is
also a drop at $k=\kout$, as can be seen in the
top panel of Figure \ref{fig:ke1}d, for example; but this drop is much less 
drastic than with
the flat spectrum extrapolation.
\label{foot:spike}
}
We integrate in time with second-order Runge-Kutta.

For the discussions of the simulations that follow, 
recall that the energies of the waves at a given 
lengthscale $\lp=1/k$ are, within unimportant multiplicative 
constants,
\begin{equation}
(\wuppl)^2\sim k^2\eupp(k) \  , \ \ (\wdownpl)^2\sim k^2\edownp(k) \ .
\end{equation}
The cascade times of the up- and down-going waves are, respectively,
\begin{equation}
\tcasup(k)\sim (k^4\edownp(k) )^{-1} \  , \ \
\tcasdown(k)\sim (k^4\eupp(k))^{-1} \ , 
\end{equation}
(see eq. [\ref{eq:cascadetime}]).
The dissipation time is 
\begin{equation}
\tdiss(k)=(\nu k^2)^{-1} \ ,
\end{equation}
(see eq. [\ref{eq:tdiss}]).
In the simulations in this section, the viscosity is 
\begin{equation}
\nu=3\cdot 10^{-5} \ ;
\label{eq:nu}
\end{equation}
this implies that $\tdiss(\kmax)=0.001$.

\subsection{Fixed Energies at the Outer Scale}
\label{sec:fixedenergy}

\begin{figure}
\plotone{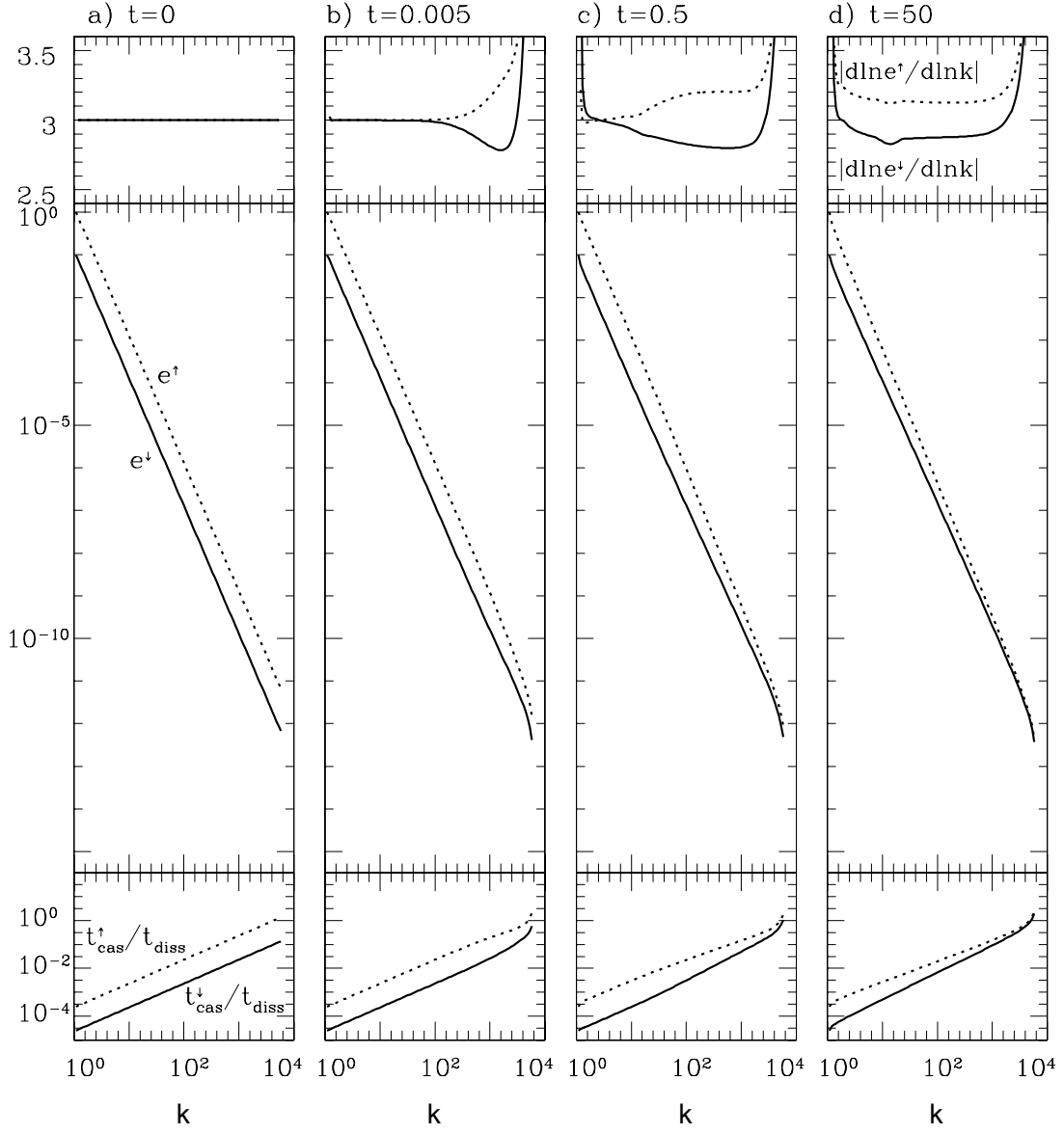}
\caption{Simulation of Kinetic Equations
with Fixed Energies at the Outer Scale
\label{fig:ke1}}
\end{figure}

For our first simulation, we fix $\eupp(\kout)= 1$ and 
$\edownp(\kout)= 0.1$ throughout the simulation.  
We do this by adding an injection term at $k=\kout$
to the right-hand side of the
kinetic equation 
(eq. 
[\ref{eq:kincode}]).
The two injection terms, $\dot{e}^\uparrow_{\rm inj}$
and $\dot{e}^\downarrow_{\rm inj}$, are adjusted
to keep $\eupp(\kout)$ and $\edownp(\kout)$ fixed.  
At the initial time, we set $\eupp\propto k^{-3}$
and $\edownp\propto k^{-3}$ (Fig. \ref{fig:ke1}a). 
These initial spectra are seemingly valid solutions of the steady-state flux
relations (eqs. [\ref{eq:epsf1}] and [\ref{eq:epsf2}]), with
$\alphaup=0$ and $\epsilon^\uparrow=\epsilon^\downarrow$.  But
the spectra are not pinned.
When they are evolved in time, it seen that they become pinned to each other at
the dissipation scale.  This pinning happens very quickly---at the 
dissipation timescale (Fig \ref{fig:ke1}b).  The reason for this pinning
was discussed in \S \ref{sec:pinned}, and can be traced in 
Figures \ref{fig:ke1}a-c.

The entire $\edownp$ spectrum adjusts to $\eupp$ on the timescale
$\tcasdown(\kout)\sim 1$.
The $\eupp$ spectrum takes longer to adjust, since 
$\tcasup(\kout)\sim 10$.
By $t=50$, steady state is reached (Fig \ref{fig:ke1}d).  
We discuss the resulting steady state spectra in 
\S \ref{sec:steadystatenumerics}.
We find that, in steady state, the energy injection rates required to keep 
the outer scale energies fixed are
\begin{equation}
{\dot e}_{\rm inj}^\uparrow = 0.136\ \ , \ \ \ 
{\dot e}_{\rm inj}^\downarrow = 0.114\ \ ;  
\label{eq:ecascade}
\end{equation}
these values are used in our second simulation.

\subsection{Fixed Fluxes}

\label{sec:fixedfluxke}

In our second simulation, we inject energy at fixed rates
at $k=\kout$, allowing $\eupp(\kout)$ and $\edownp(\kout)$
free to evolve.
We use the same injection rates as we
found in steady state in the previous simulation
(eq. [\ref{eq:ecascade}]).
We add to $\eupp(\kout)$ and $\edownp(\kout)$ at these fixed rates throughout
the present simulation.
Since we add to $\eupp(\kout)$ and $\edownp(\kout)$ at the same
rate as we did in steady state in the previous simulation, we
expect that our second simulation will reach the same steady
state as did the first one.  We shall show that it does.
Note that, strictly speaking, ${\dot e}_{\rm inj}^\uparrow$
and $\dot{e}_{\rm inj}^\downarrow$ are not precisely 
equal to the fluxes.
Nonetheless, we ignore this subtlety and refer
to the present simulation as one of constant fluxes.
We shall evaluate the fluxes more precisely below.

\begin{figure}
\plotone{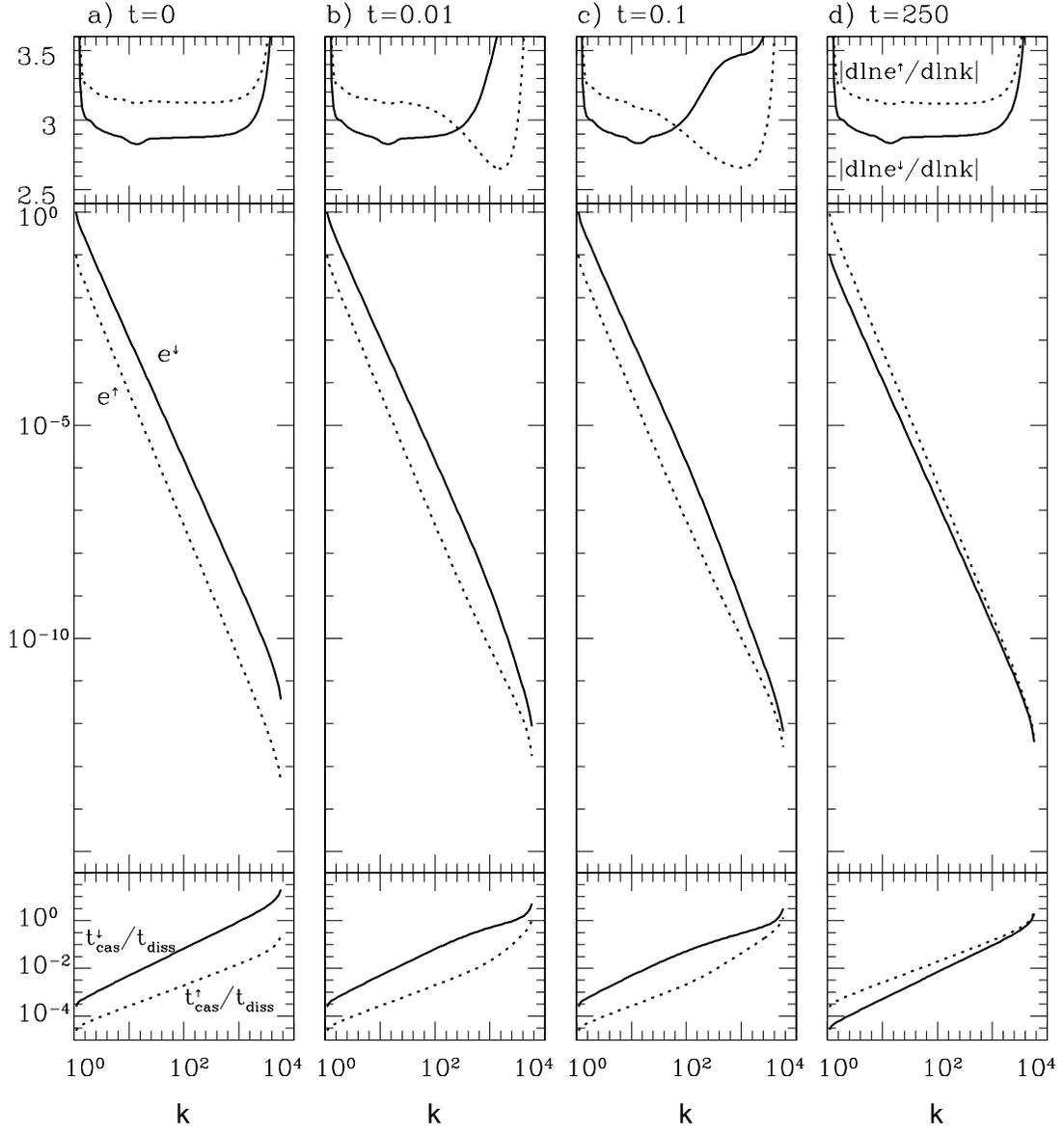}
\caption{
Simulation of Kinetic Equations with
Fixed Fluxes
\label{fig:ke2}
}
\end{figure}

For our initial condition, we use the spectra obtained 
in steady state in the first simulation, but 
multiplied by a constant; specifically,
$\eupp(k)\rightarrow\eupp(k)/10$ and 
$\edownp(k)\rightarrow\edownp(k)\cdot 10$ (see Fig. \ref{fig:ke2}a).
With these initial spectra, and with the aforementioned 
fluxes,
the flux 
relations (eqs. [\ref{eq:epsf1}] and [\ref{eq:epsf2}]) are
satisfied---since they were satisfied before
the multiplication and division by 10. But, once again, this is not a valid
steady state solution because the spectra are not pinned at 
the dissipation scale.  When the spectra are evolved
in time, the spectra are first pinned (Figs. \ref{fig:ke2}a-c).
In this simulation, the spectra are initially 
pinned where $\tcasdown\sim\tdiss$,
which is
at $k\sim 1,000$.  Therefore the pinning timescale is 
$\sim \tdiss(k=1,000)\sim 0.03$.  

As expected, the same steady state
is reached as in the first simulation (Fig. \ref{fig:ke2}d).
In reaching steady state, the two spectra must cross. 
This happens at $t\sim 40$, when the two spectra are 
nearly identical, with 
slopes equal to $-3$, and with outer scale energy equal to $\sim 0.3$.
The time to reach steady state 
is considerably longer in the constant flux simulation than
in the constant energy one.
Steady state is reached at   
$t\sim 250$, 
as opposed to $t\sim 50$ in the previous simulation. 

\subsection{Energies and Fluxes in Steady State}
\label{sec:steadystatenumerics}

We can quantitatively
compare the behaviour in steady state with our calculations
in \S \ref{sec:sses}.  We consider the steady state 
reached in the two simulations discussed above.  Recall that
both simulations reach the same steady state, at which point
the spectra are 
as shown in Figures \ref{fig:ke1}d and \ref{fig:ke2}d.
For our comparisons, we need the following quantities:
$\nu=3\cdot 10^{-5}$ (eq. [\ref{eq:nu}]), 
$\kout=1.09$ (eq. [\ref{eq:outer}]),
$\lpar/v_A=1/2$ (eq. [\ref{eq:lpar}]),
$\eupp(\kout)=0.63$, and $\edownp(\kout)=0.066$.\footnote{
For our first simulation, we 
actually fixed $\eupp(\kout)=1$ and $\edownp(\kout)=0.1$.
However, 
as we discuss in footnote 
\ref{foot:spike}, there is a drop in energy between the first
and second grid point.
The values $\eupp(\kout)=0.63$, and $\edownp(\kout)=0.066$
are found by taking the power laws seen in steady state for $\eupp(k>\kout)$
and $\edownp(k>\kout)$, and extrapolating them to $k=\kout$.
\label{foot:amplitude}
}

From equation (\ref{eq:ldiss}), the dissipation
wavenumber is 
\begin{equation}
\kdiss={\lpar\over \nu v_A}(\kout^6 \eupp(\kout)\edownp(\kout))^{1/2} =
4,400 \ \ \ \ ,
\label{eq:kdisspred}
\end{equation}
in agreement with 
the value seen in Figure \ref{fig:ke1}d.  

Equation (\ref{eq:alphaup}) gives 
\begin{equation}
\alpha={\ln [ \eupp(\kout)/\edownp(\kout)]
\over
\ln [\kout^4\eupp(\kout)\edownp(\kout)(\lpar/\nu v_A)^2] }=0.14 \ ,
\label{eq:alphanum}
\end{equation}
which implies that the steady state slopes should be
\begin{eqnarray}
-{d\ln\eupp\over d\ln k}&=&3+\alphaup=3.14 \\ 
-{d\ln\edownp\over d\ln k}&=&3-\alphaup=2.86 \ .
\end{eqnarray}
This is in agreement with the top panel of Figure \ref{fig:ke1}d, 
which shows that the slope of $\eupp$ in the inertial range,
while varying slightly with $k$, mostly remains between
3.1 and 3.15; the slope of
$\eupp$ is between 2.85 and 2.9.

\begin{figure}
\plotone{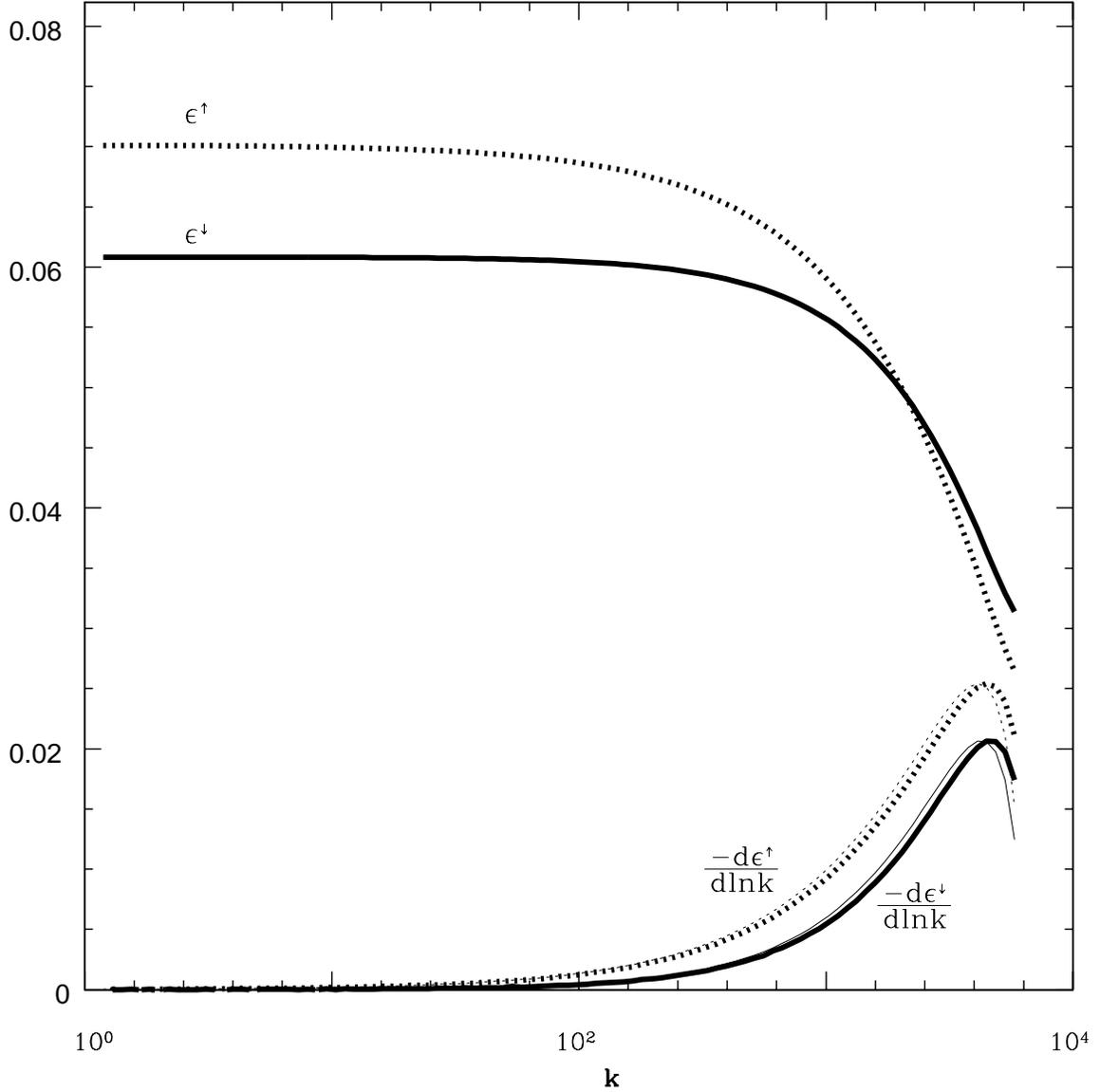}
\caption{Steady State Fluxes and Flux Gradients
\label{fig:kefluxes}
}
Fluxes and flux gradients are shown as bold solid and
dotted lines.  Also 
shown as unlabelled thin lines that nearly overlap the flux
gradients are $\nu k^4\eupp$ (thin dotted line)
and $\nu k^4\edownp$ (thin solid line).  
The spectra from which the fluxes were extracted 
are shown in
Figure \ref{fig:ke1}d. 
\end{figure}

To derive the steady-state formulae that we have used thus
far in this section, heuristic arguments suffice.
However, to relate energy spectra to fluxes, the kinetic
equation is required.
From equation (\ref{eq:epsprod}), we should have
\begin{equation}
\epsilon^\uparrow
\epsilon^\downarrow=[1.87\cdot \kout^6\eupp(\kout)\edownp(\kout)(\lpar/v_A)]^{2}
=0.0043 \ ,
\label{eq:epspred1}
\end{equation}
where the numerical factor 1.87 is obtained from the kinetic equation.
A more important application of the kinetic equation is to derive
the ratio of the fluxes, for which heuristic arguments are
useless.  From  
equation (\ref{eq:alphaupapprox}),
we should have 
\begin{equation}
\epsilon^\uparrow/\epsilon^\downarrow-1=\alphaup=0.14 \ ,
\label{eq:epspred}
\end{equation}
when we use the value of $\alphaup$ predicted in equation (\ref{eq:alphanum}).
To compare the above predictions for the fluxes
(eqs. [\ref{eq:epspred1}] and [\ref{eq:epspred}])
with the fluxes seen in the numerical simulation, we plot the
latter in Figure \ref{fig:kefluxes}. 
We calculated $\epsilon^\uparrow$ 
with equation (\ref{eq:flux}), using the steady
state spectra
shown in Figure \ref{fig:ke1}d. 
The quantity $S^\downarrow(p,q)$ that appears
in equation (\ref{eq:flux}) is given in equation (\ref{eq:s}).
Since the kinetic code 
calculates $S^\downarrow(p,q)$ in order to evolve the
kinetic equations (see eq. [\ref{eq:kincode}]), 
it is trivial to modify
the code so that it can be used to evaluate the flux. 
To calculate $\epsilon^\downarrow$, we used the
analogue of equation (\ref{eq:flux}), with $\uparrow$
and $\downarrow$ switched.
We see in Figure \ref{fig:kefluxes} that the fluxes are
nearly independent of $k$ in the inertial range, 
with $\epsilon^\uparrow\simeq 0.070$ 
and $\epsilon^\downarrow\simeq 0.061$.
These values give 
$\epsilon^\uparrow/\epsilon^\downarrow-1\simeq 0.15$, 
in agreement with equation (\ref{eq:epspred}).
They also give $\epsilon^\uparrow\epsilon^\downarrow\simeq 0.0043$,
in agreement with equation (\ref{eq:epspred1}).

From Figure \ref{fig:kefluxes}, dissipation affects
$\epsilon^\uparrow$ and $\epsilon^\downarrow$ for $k\gtrsim 100$.
Thus dissipation has a big reach. 
If we add the dissipation term to the flux-conservation form
of the kinetic equation (eq. [\ref{eq:kineticflux}]), 
then we see that in steady state the kinetic equation must satisfy
\begin{equation}
-{d\epsilon^\uparrow\over d\ln k}=\nu k^4 \eupp \ .
\label{eq:sske}
\end{equation} 
The left-hand side of this equation gives 
the rate of energy increase (per logarithmic band in $k$) 
due to the $k$-space
divergence of the flux; in steady state, it
must be balanced by the energy loss rate
due to dissipation.  In Figure \ref{fig:kefluxes}, we plot
$-(d/d \ln k)\epsilon^\uparrow$ as a bold dotted line;
we evaluated it by taking the numerical derivative of the
displayed $\epsilon^\uparrow$.  The quantity
$\nu k^4 \eupp$ is the unlabelled thin dotted line that
nearly overlaps $-(d/d \ln k)\epsilon^\uparrow$.
Thus equation (\ref{eq:sske}) is satisfied in the simulation.
The same is true for the down waves, which are shown as solid lines
in  Figure \ref{fig:kefluxes}.

\subsection{Decaying Turbulence}

\begin{figure}
\plotone{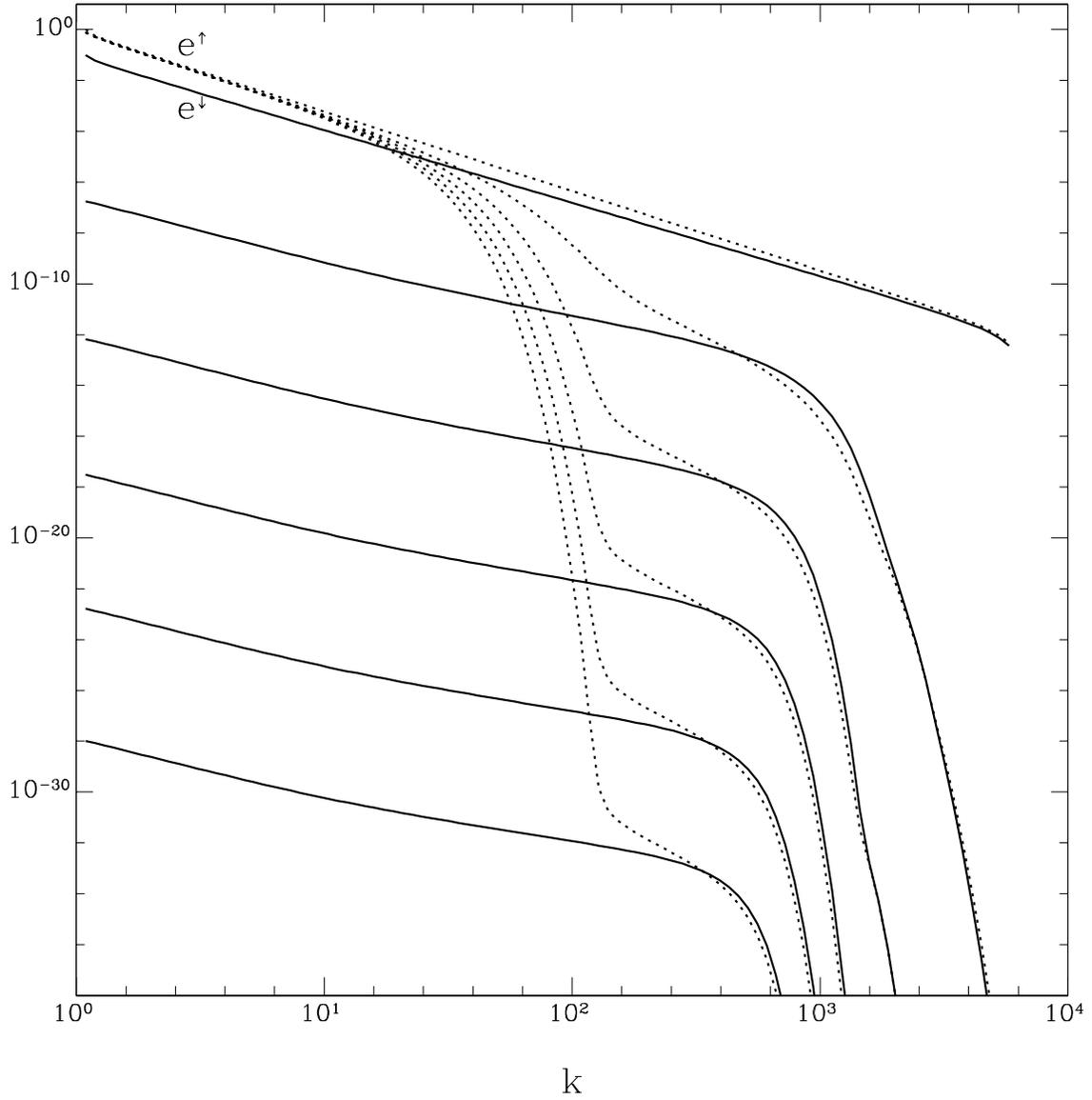}
\caption{
Decaying Simulation of Kinetic Equations}
Spectra of $\eupp$ (dotted lines) and $\edownp$
(solid lines) at the following times:
$t=0, 25, 50, 75, 100, 125$ (from top to bottom).
\label{fig:decay}
\end{figure}

For our third simulation, we allow the spectra to decay without
injecting any energy.  The initial condition
is the steady state spectra from the previous simulation, 
see Figure \ref{fig:decay}.  
At the outer scale, the cascade time of the down-waves
is $\sim 1/\eupp(\kout)\sim 1$, so $\edownp(\kout)$ 
decays on this timescale.  
More precisely, from the values
plotted in Figure \ref{fig:decay}, $\edownp\propto \exp(-0.5 t)$
at fixed $k$.

The cascade time of the up-waves is much longer.
Initially, $\tcasup\sim 10$ at the outer scale; 
as $\edownp$ decays, 
$\tcasup$ increases exponentially.  So
$\eupp(\kout)$ does not evolve.  At larger $k$'s,
$\eupp$ is cut off by dissipation; the
dissipation wavenumber is 
$\sim (\nu t)^{-1/2}\sim 200 t^{-1/2}$.

The spectra appear to evolve in a self-similar manner, with
the spectra remaining pinned at their dissipation scale.
The end result is that the energy in the down-waves disappears,
while the energy in the up-waves is nearly unchanged.  
Decaying weak turbulence is unstable: an initial imbalance
between up and down waves is magnified exponentially.
A similar instability occurs in strong turbulence, as suggested by
Dobrowolny et al. (1980) in the context of the solar
wind, and as seen in numerical simulations of
strong turbulence
(Maron \& Goldreich 2001, and
Cho et al. 2002).

\section{A Model of the Kinetic Equations: Coupled Diffusion Equations}

It is instructive to 
model the kinetic equations with
two coupled diffusion equations.
Investigation of these model equations illustrates
that the pinning
of spectra is quite general; the only requirement is that
the cascade time 
of one type of wave be inversely related to the energy
of the other type.  A second reason for considering model
equations is that they
can be simulated much faster and more easily than the
kinetic equations.
Finally, by contrasting the kinetic equations
with the model equations, we can gain insight into
the behaviour of the kinetic equations.

\subsection{Derivation of Coupled Diffusion Equations}

Our model equation is
\begin{equation}
\frac\partial{\partial t}
k\eupp=
-{\partial\over\partial k}  \epsupt \ ,
\label{eq:diffeq}
\end{equation}
where $ke^\uparrow$ is the energy per unit $k$,
and $\epsupt$ is the flux.  This equation is 
similar in form to equation (\ref{eq:kineticflux}).  
Recall from \S \ref{sec:energyandflux}
that we define the flux as the energy flow rate across
a fixed $k\equiv\vert\bk\vert$.  It is the ``1-D flux,''
since we average over angles;
similarly, $ke^\uparrow$ is the ``1-D energy density.''
For $\epsupt$, we choose a form that depends on the
following quantities evaluated at $k$:
$\edownp$, $\eupp$, and $\partial_k\eupp$.
By contrast, the kinetic equations depend on $\eupp$ and $\edownp$
evaluated at a range of wavenumbers, so 
unlike the kinetic equations, the diffusion equations
are exactly local in $k$.
For the cascade time of the energy in the
up waves to have the 
correct form (eq. [\ref{eq:cascadetime}]),
i.e., $\tcasup=
(\lp/\wdownpl)^2
(v_A/\lpar)\sim (k^4\edownp)^{-1}(v_A/\lpar)$, 
we choose
\begin{equation}
\epsupt=-{\lpar\over v_A}\edownp k^{6-\beta}{\partial\over\partial k}
\Big( k^{\beta+1} \eupp \Big) \ , 
\label{eq:diffflux}
\end{equation}
where $\beta$ remains to be specified.
Of course the above relation for the flux gives the correct
steady-state scaling:  $\eupp\edownp\propto k^{-6}$. 
The evolution equations for $\edownp$ are the same as
for $\eupp$ (eqs.
[\ref{eq:diffeq}] and [\ref{eq:diffflux}]), 
with
$\uparrow$'s and $\downarrow$'s interchanged.  

With $\eupp\propto k^{-(3+\alphaup)}$, 
$\edownp\propto k^{-(3-\alphaup)}$,
we can relate the steady-state fluxes to the energies:
\begin{eqnarray}
\epsupt&=&\tilde{f}(\alphaup)
[k^6\eupp(k)\edownp(k)]
\frac{\lpar}
{v_A}  \ ,  \label{eq:epsft1} \\ 
\epsdownt&=&\tilde{f}(-\alphaup)
[k^6\eupp(k)\edownp(k)]
\frac{\lpar}
{v_A}  \ , \label{eq:epsft2}
\end{eqnarray}
where  
\begin{equation}
\tilde{f}(\alphaup)=2+\alphaup-\beta \ .
\end{equation}
These relations are analogous to equations (\ref{eq:epsf1}) and (\ref{eq:epsf2}).
To make the analogy closer,
we choose $\beta$ so that $\tilde{f}$ has the same dependence on $\alphaup$
as does $f$ in the limit that the cascade is nearly balanced 
(i.e., $\vert\alphaup\vert\ll 1$).
Since $\epsupt/\epsdownt=\tilde{f}(\alphaup)/\tilde{f}(-\alphaup)\simeq
1+\alphaup/(1-\beta/2)$ in this limit,
we see by comparison with equation (\ref{eq:alphaupapprox}) that 
we should choose $\beta=0$.

Collecting results, our model
equations are two coupled diffusion equations:
\begin{eqnarray}
{\partial \over \partial t}
\eupp
&=&
{\lpar\over v_A}{1\over k}{\partial \over \partial k} \Big (\edownp k^6
{\partial \over \partial k} k
\eupp \Big )
-\nu k^2 \eupp \ , \label{eq:diffeq1} \\  
{\partial \over \partial t}
\edownp 
&=&
{\lpar\over v_A}{1\over k}{\partial \over \partial k} \Big (\eupp k^6
{\partial \over \partial k} k
\edownp \Big )
-\nu k^2 \edownp  \ , \label{eq:diffeq2}
\end{eqnarray}
after including dissipative terms.

These equations, while much simpler than the kinetic equations,
share many of the same features.  Since the cascade time of
the up-going waves is inversely proportional to the energy of
the down-going waves (and vice versa), the argument for the 
pinning of the spectra in weak turbulence (see \S
\ref{sec:pinned}) applies here as well.
Moreover, in steady state, these diffusion equations suffer
from the same degeneracy as does the kinetic equation:
the constancy of $\epsupt$ and
$\epsdownt$ is insufficient to determine the scaling of
$\eupp$ or $\edownp$ separately.
This degeneracy is partially broken by the dependence of the fluxes
on the slopes (eqs. [\ref{eq:epsft1}] and [\ref{eq:epsft2}]);
we have chosen this dependence so that it is the same as
for the kinetic equations in the limit that the cascade
is nearly balanced ($\vert\alphaup\vert\ll 1$).
In the following section, we present numerical simulations of these
equations.  

\subsection{Numerical Simulations of Coupled Diffusion Equations}

We run two simulations of the coupled diffusion equations.
As with the kinetic simulations described in \S \ref{sec:numkin},
the first simulation has fixed energy at the outer scale, and
the second simulation has fixed flux.
And, as before, $2\lpar/v_A$ is set to unity,
functions of $k$ are evaluated at
$k=2^{i/8}$, $i=1,...,100$, and the viscosity is 
$\nu=3\cdot 10^{-5}$. 

\subsubsection{Fixed Energies at the Outer Scale}

\begin{figure}
\plotone{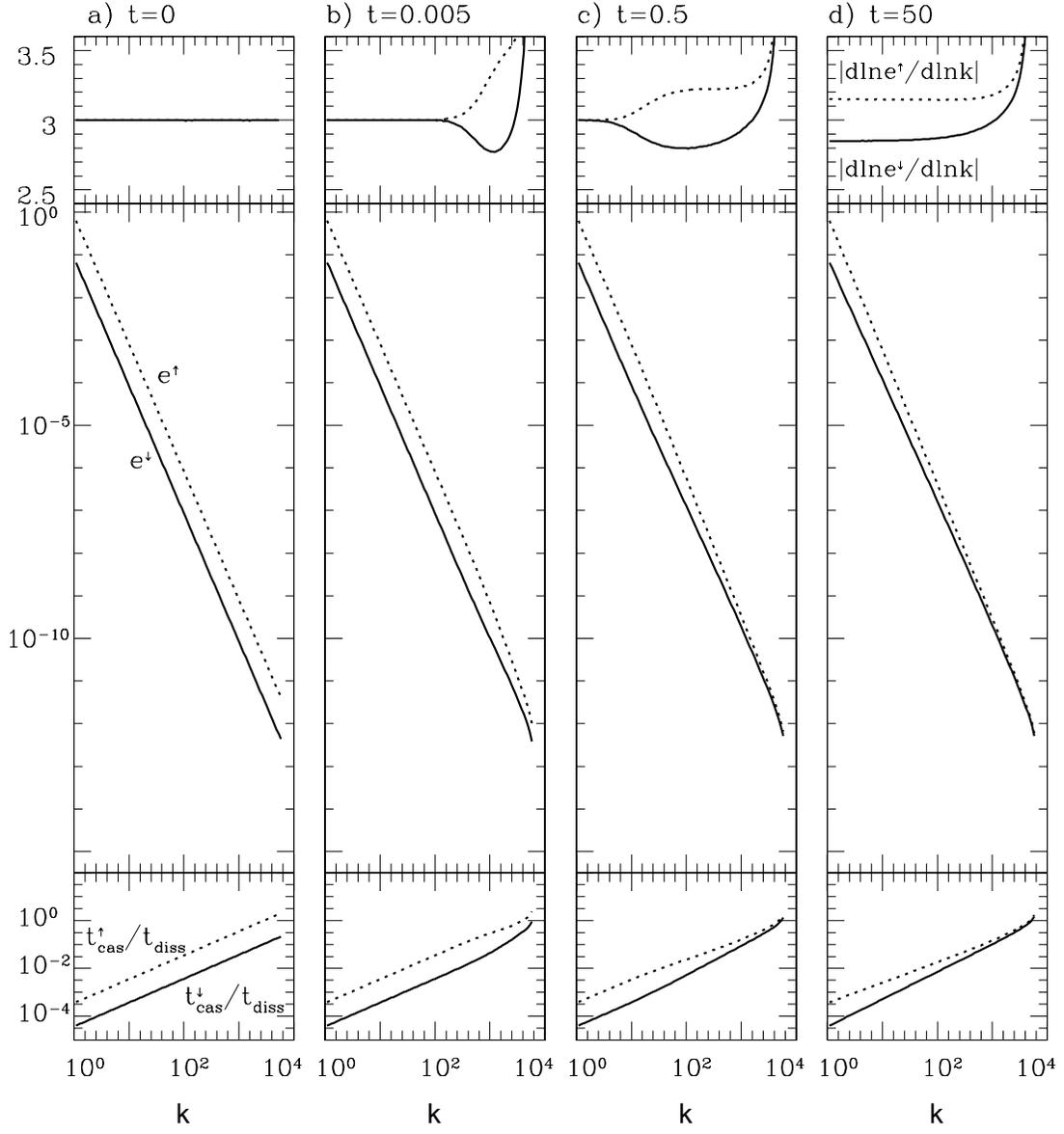}
\caption{
Simulation of Diffusion Equations with 
Fixed Energies at the Outer Scale
\label{fig:diff1}
}
\end{figure}

For our first simulation, we fix $\eupp(\kout)=0.63$ and
$\edownp(\kout)=0.066$.\footnote{We choose these numbers
so that the spectra will have the same amplitude here
as in the kinetic simulation; see footnote \ref{foot:amplitude}.}
The evolution is shown in
Figure \ref{fig:diff1}.  It is very similar to the evolution
of the kinetic simulation (Fig. \ref{fig:ke1}).  As before,
the spectra are pinned quickly.  And since, by design, the 
predicted steady
state relations for the diffusion equations are the
same as those for the kinetic equations 
(eqs. [\ref{eq:kdisspred}]--[\ref{eq:epspred}]),
the steady state behaviour of the two simulations are nearly
identical; compare Figures \ref{fig:diff1}d and \ref{fig:ke1}d.
There are two differences worthy of note.
First, 
the slopes of $\eupp$ and $\edownp$ do not have a spike
near $k=\kout$.  The presence of such a spike in the kinetic simulation
is due to the extrapolation of the spectra to $k<\kout$ (see footnote
\ref{foot:spike}).  Since the diffusion equations are exactly
local in $k$, such an extrapolation is unnecessary, and so the
behaviour is much smoother near $k=\kout$.
Second, in steady state, the diffusion simulations
yield spectral slopes that change more gradually 
as a function of $k$ near the dissipation scale;
compare 
the top panel of Figure \ref{fig:ke1}d with that of 
Figure \ref{fig:diff1}d.

\subsubsection{Fixed Fluxes}
\begin{figure}
\plotone{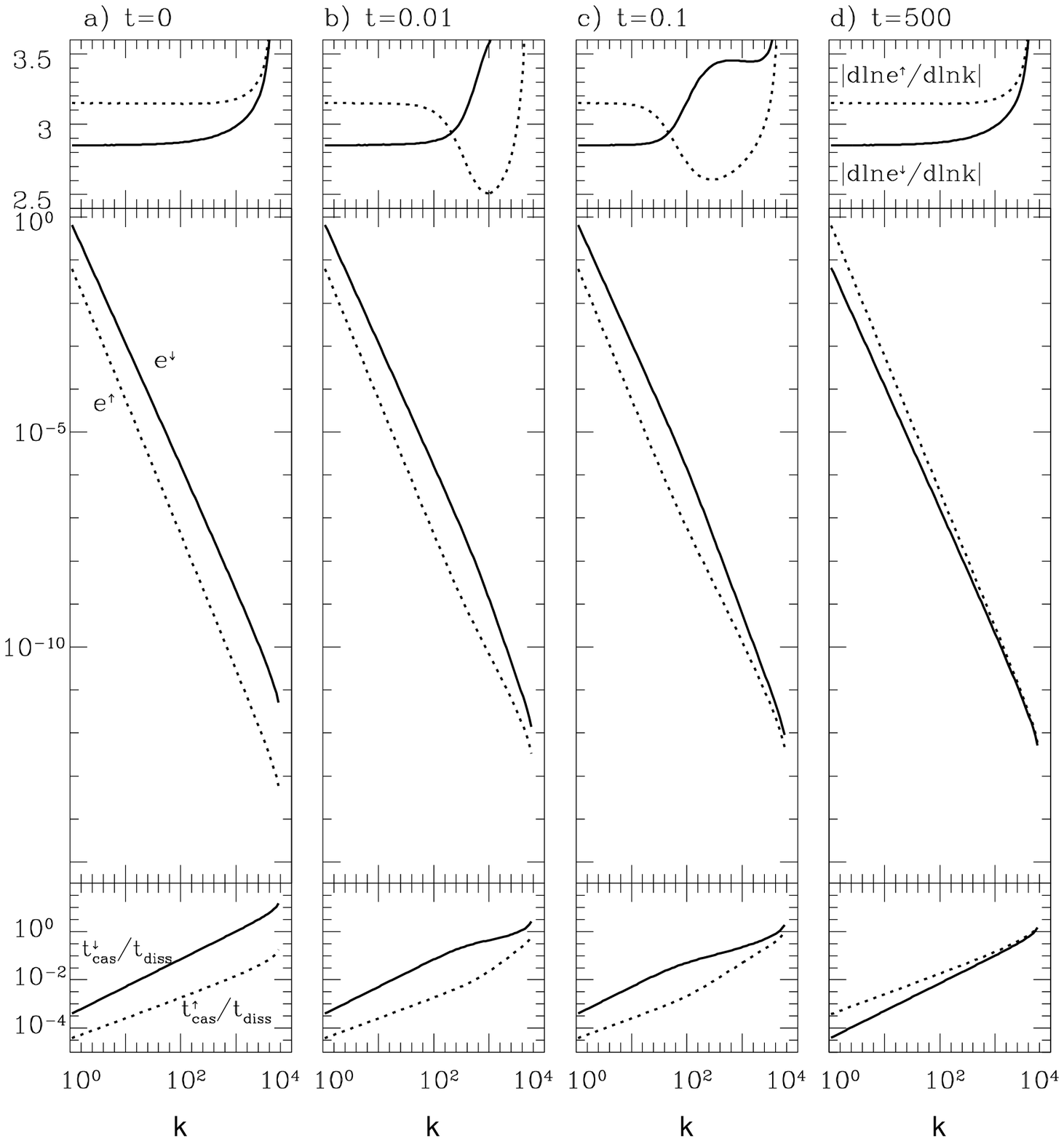}
\caption{
Simulation of Diffusion Equations with
Fixed Fluxes
\label{fig:diff2}
}
\end{figure}

For the second simulation, we inject energy at $k=\kout$ at a fixed rate,
equal to that found in steady state in the previous simulation.
For the initial condition, we divide the previously found steady state
value of $\eupp$ by 10, and we multiply $\edownp$ by 10. 
The evolution is shown in Figure \ref{fig:diff2}.  It is very similar
to that found in the corresponding kinetic simulation
(Figure \ref{fig:ke2}).

\section{Large Magnetic Prandtl Number}
\label{sec:prandtl}

In incompressible MHD,
diffusive dissipation of kinetic and magnetic
energies is accounted for by addition of the terms
$\nu\nabla^2 \bov$ and $\eta\nabla^2\bB$ 
to the right-hand sides of
equations (\ref{eq:eomv}) and (\ref{eq:eomb}), respectively, 
where $\nu$ is the viscosity and $\eta$ is the resistivity.
The magnetic Prandtl number is defined as 
\begin{equation}
{\rm Pr}\equiv\nu/\eta \ .
\end{equation}
Until now, we have assumed that Pr$=1$.  In this section,
we examine weak MHD turbulence when Pr$\gg 1$.\footnote{
Strong MHD turbulence with Pr$\gg 1$ occurs in many astrophysical
settings.}

With arbitrary $\nu$ and $\eta$, the equation for 
$\bwup$ (eq. [\ref{eq:eom1}]) becomes, 
\begin{equation}
\pt \bwup + v_A \pz\bwup=-\bwdown\bcdot\bnabla\bwup-\bnabla P 
+{1\over 2}(\nu+\eta)\nabla^2\bwup+{1\over 2}(\nu-\eta)\nabla^2\bwdown \ 
\ ;
\end{equation}
the equation for $\bwdown$ is the same but with $\uparrow$
and $\downarrow$ interchanged, and $v_A\rightarrow -v_A$.
To derive the kinetic equation for $\eupp$, we first Fourier
transform the above equation; the transformed dissipation terms 
are $-(k^2/2)(\nu+\eta){\bld w^\uparrow_k}-
(k^2/2)(\nu-\eta){\bld w^\downarrow_k}$.
When we take the dot product of the Fourier-transformed equation
with ${\bld w^\uparrow_k}$, and then take the expectation value of the
result, we can neglect
the dissipation term proportional to 
$\left<{\bld w^\uparrow_k}\bcdot{\bld w^\downarrow_k}\right>$,
since it is much smaller than the term proportional 
to $\left<({\bld w^\uparrow_k})^2\right>$.  This is because
in upgoing waves at fixed 
$\zup\equiv z-v_A t$ are uncorrelated with the downgoing ones
(see \S 
\ref{sec:prelim}).
As a result, the kinetic equation for high magnetic Prandtl 
number is the same as that for
Pr$=1$ (eq. [\ref{eq:kineticequation}]), but with $\nu k^2 \eupp$ 
replaced by $(\nu+\eta)k^2\eupp/2\simeq \nu k^2\eupp/2$. 
This might appear to be a surprising result.  
In the limit of vanishing $\eta$, the energy spectra 
are 
cut off below the viscous scale, i.e., the scale set by $\nu$
(eq. [\ref{eq:ldiss}], within a factor of 2).
One might have expected that the viscous scale should only affect
the kinetic energy spectrum.  Yet the magnetic energy spectrum
is also cut off below the viscous scale.
This can be understood as follows: upgoing waves each have
equal magnetic and kinetic energies, and similarly for downgoing waves.  
As an up-wave with lengthscale slightly larger than the viscous scale
is gradually cascaded to sub-viscous scales, 
its kinetic energy is dissipated by viscosity.  Since the cascade
time in weak turbulence is much longer than the waveperiod, as the
up-wave's kinetic energy is dissipated, its magnetic energy is
converted into kinetic energy so that its magnetic and kinetic energies
can remain nearly equal.  As a result, both kinetic and magnetic energies are
dissipated at the viscous scale.

\section{Hyperviscosity and the Bottleneck Effect}
\label{sec:bottle}

In many different types of turbulent cascades, the
energy spectrum exhibits a hump on scales slightly larger
than the dissipation scale (e.g., Falkovich 1994).
The hump is 
particularly pronounced in numerical
simulations 
that use ``hyperviscosity,''
a trick whereby the diffusive term
$\nu k^2 \bld{v}$ is replaced by
$\nu_{n}k^n \bld{v}$, where $n$ is typically 4 or 8, and
$\nu_n$ is the hyperviscosity; the resistivity is modified
in a similar manner 
(e.g., Borue \& Orszag 1995, Biskamp 2000).
With this trick, a smaller part of the spectrum is
affected by dissipation, 
so a longer inertial range can be simulated with
a fixed resolution.
Although this trick does work, 
there is a problem: the spectrum on scales slightly larger than
the dissipation scales is made flatter. 
The energy, in effect, is backed up.  This
is the bottleneck effect.  
It can be particularly problematic in 
simulations of strong MHD turbulence,
where the energy backup can affect lengthscales considerably
larger than the dissipative scales (Biskamp \& 
M\"uller 2000). 

\begin{figure}
\plotone{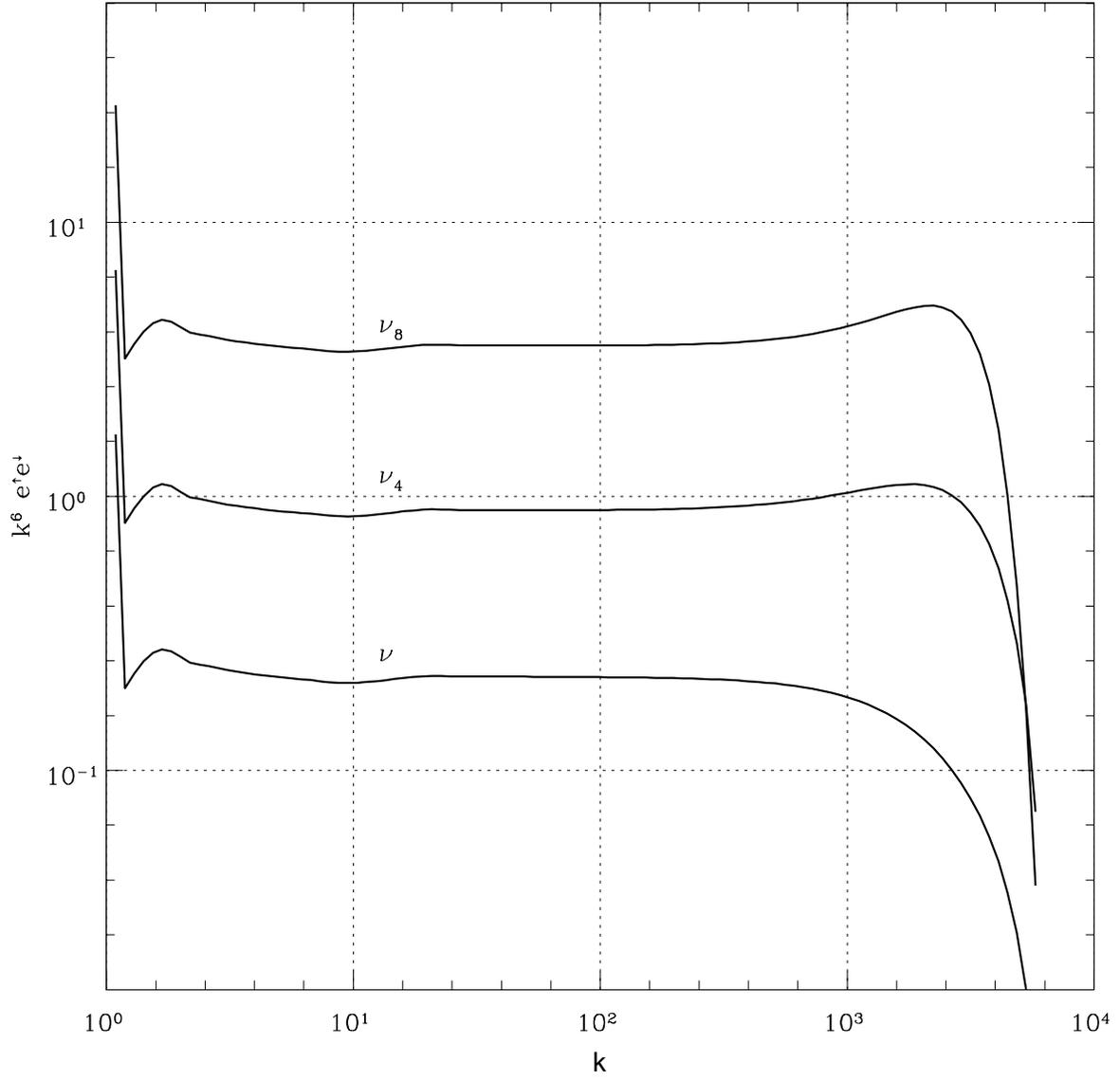}
\caption{
The Bottleneck Effect
\label{fig:bottle}
}
Three spectra, offset for clarity; the top spectrum has viscous and resistive
terms $\nu_8 k^8$, the middle has $\nu_4 k^4$, and the bottom has $\nu k^2$. 
\end{figure}

This motivates us to investigate the bottleneck effect
in weak turbulence.  
If the bottleneck effect appears,
its interpretation will be simpler than in 
strong turbulence.
We perform two simulations of the kinetic equation, one
with hyperviscosity of the form $\nu_4 k^4$, and the other with
$\nu_8 k^8$.
In each simulation, we fix the energies at the outer scale,
$\eupp(\kout)=1$, $\edownp(\kout)=0.1$, and allow 
the spectra to reach steady state.  The steady state
spectra of $k^6\eupp\edownp$ are shown in 
Figure \ref{fig:bottle}, offset for clarity.
Also shown is the simulation with ordinary viscosity
described in \S  \ref{sec:fixedenergy}.
From this figure it is apparent that weak turbulence
suffers from the bottleneck effect; the effect
becomes larger with increasing hyperviscous exponent.

The bottleneck effect can be understood as follows.
Consider an up-wave with a lengthscale slightly larger
than the dissipation scale.  It is cascaded by down-waves
that have slightly different lengthscales than its own.
Hyperviscosity gives a sharper dissipative cutoff to 
the down-wave spectrum than ordinary viscosity.  
Therefore a hyperviscous simulation has fewer down-waves
in the vicinity of the dissipation scale, 
and the cascade time of the up-wave is longer.
For the up-wave energy flux to be independent of lengthscale,
a longer cascade time implies a larger energy.  As a result,
the spectrum is flatter on scales slightly larger than 
dissipative ones.
Falkovich (1994) offers a similar explanation for the bottleneck effect
in hydro turbulence.

\section{Discussion}

In this paper we describe imbalanced weak turbulence
and solve the steady state cascade.
In a future paper, we will extend the result
to the strong cascade.  One of our ultimate goals is to
develop a theory of imbalanced strong turbulence to 
apply to the solar wind, where imbalance is
observed.

Although strong turbulence is more generally applicable than weak,
the latter is a simple and illuminating model.
There are a number of issues in strong turbulence that are
not understood.   Weak turbulence can be used as the first
step in explaining them.  For example, in this paper we
examined the effect of a large magnetic Prandtl number on weak MHD
turbulence.  We also
found that the bottleneck effect appears in weak turbulence,
where its interpretation is straightforward.
There are a number of
other issues that we intend to investigate in weak turbulence
as a prelude to understanding them in strong turbulence; for example,
reconnection and the turbulent dynamo.

\acknowledgments
Research reported in this paper was supported by 
NSF grant AST-0098301.

\appendix
\section{Appendix: Kinetic Equation in Weak Turbulence}
\subsection{Preliminaries}
\label{sec:prelim}
Kinetic equations describing weak MHD turbulence
are derived by Galtier et al. (2000, 2002). 
In this appendix, we present a more
physical derivation. We compare the two derivations
in \S \ref{sec:compare}.

We consider the evolution of $\bwupp$
in a plane that is transverse to $\zhat$ and 
moving with velocity
$v_A\zhat$, i.e., with fixed $\zup\equiv z-v_A t$. 
Recall that $\zhat$ is in the direction of the mean magnetic field.
Changing variables
from $z$ to $\zup$ in equation (\ref{eq:eomrmhd1}) gives
\begin{equation}
{\partial\over\partial t}\Big\vert_{x,y,\zup}
\bwupp=-\bwdownp\bcdot\bnablap\bwupp
+\bnablap\nabla_\perp^{-2}(\bnablap\bwdownp{\bld :}\bnablap\bwupp) \ .
 \label{eq:eomrmhdappendix}
\end{equation}
In weak turbulence, the
parallel length over which disturbances are correlated, $\lpar$, 
is small; in particular, in the time that an up-wave crosses
a down-going wavepacket, $\Delta t\sim\lpar/v_A$, its distortion is
less than unity: 
$\Delta \wuppl/\wuppl\sim (\wdownpl/v_A)\lpar/\lp\ll 1$ 
(eq. [\ref{eq:inter}]).\footnote{We neglect the factor of 2 associated
with the fact that the relative velocity of up and down waves is $2 v_A$.}
Therefore, $\bwupp$ undergoes small uncorrelated changes each 
$\Delta t\sim\lpar/v_A$, and its evolution is analogous to a random walk,
with step-size $\Delta\wuppl\sim\wuppl (\wdownpl/v_A)\lpar/\lp\ll 1$.
Our goal in this appendix is to quantify the random walk; the resulting
equation is the kinetic equation.
The kinetic equation
is only valid when the turbulence is weak: its
derivation hinges on the assumption 
that the time for up-waves to cascade
is longer than the correlation time of down-waves
at fixed $\zup$, i.e., that $(\wdownpl/v_A)\lpar/\lp\ll 1$.

Equation (\ref{eq:eomrmhdappendix}) is linear in
$\bwupp$.  
Nonlinearity arises 
because $\bwupp$ modifies $\bwdownp$ 
(through eq. [\ref{eq:eomrmhd3}]); this modification backreacts
on $\bwupp$ through equation (\ref{eq:eomrmhdappendix}).
Nonetheless, 
this backreaction can be neglected when deriving the
kinetic equation.  The reason is as follows.
Every $\Delta t\sim
\lpar/v_A$, the backreaction changes $\wuppl$ by
$\sim\wuppl [(\wdownpl/v_A)\lpar/\lp][(\wuppl/v_A)\lpar/\lp]$,
which is smaller than $\Delta \wuppl$ by the small factor
$(\wuppl/v_A)\lpar/\lp\ll 1$.  Furthermore, since this backreaction
is uncorrelated on timescales larger than $\lpar/v_A$, it only effects
a small change in the step-size of the random walk.

Therefore, in deriving the kinetic equation, we interpret equation 
(\ref{eq:eomrmhdappendix}) as a linear equation for $\bwupp$
(at fixed $\zup$); $\bwdownp$ at fixed $\zup$ 
is viewed as a function with known statistical properties.
To simplify our derivation, we first solve a model problem:
the linear random oscillator.
The
extension to weak turbulence will then be straightforward.

\subsection{A Toy Problem: the Linear Random Oscillator}
\label{sec:toy}
We consider the evolution of a simple random oscillator $\psi$:
\begin{equation}
{d\over {dt}} \psi(t)=iA(t)\psi(t) \ ,  \label{eq:randosc}
\end{equation}
where $\psi$ is a complex scalar and
$A$ is a real random variable; the factor $i$ ensures that 
the energy $\vert \psi(t)\vert^2$ is conserved.\footnote{
For a textbook discussion of the linear random 
oscillator, see van Kampen (1992).}
Our goal is to
calculate
the evolution of $\left<\psi(t)\right>$,
where
angled brackets denote an average
over an ensemble of $A$'s, not over time.
We assume that the values of $A$ 
at two different times are statistically independent of each other
whenever the two times are separated by
more than the correlation time, $\taucor$; 
we make this more precise in footnote \ref{foot:corr}.
For simplicity, we take $A$ to have zero mean,
$\left<A(t)\right>=0$;
the extension to $A$ with non-zero mean is trivial.
We assume that the statistical properties
of $A$---such as $\taucor$ 
and $\arms\equiv \left<A^2\right>^{1/2}$---vary
only on timescales
much larger than $\taucor$.   
Note that $\psi$ corresponds to $\wuppl$ in equation
(\ref{eq:eomrmhdappendix}); $A$ corresponds to $\wdownpl$,
or, more specifically, to $\wdownpl/\lp$; and
$\taucor$ corresponds to $\lpar/v_A$.

There 
are two limiting regimes for the random oscillator, depending
on whether $\arms\taucor$ is less than or greater than unity.
In the following,
we take $\arms\taucor\ll 1$; this regime 
corresponds to weak turbulence. 
The change in $\psi$ within the time 
$\taucor$ is of order
$\arms \taucor \psi$, which is much smaller than $\psi$.
Thus the correlation time of $A$ is smaller than 
the ``cascade time'' of $\psi$.
Since $A$ is uncorrelated on timescales larger than 
$\taucor$, $\psi$ undergoes small uncorrelated changes each
$\taucor$, and thus its long-time evolution is a random
walk.
Intuitively, we expect that the time evolution 
of the statistical properties of $\psi$---in particular, $\left<\psi\right>$--- 
can be represented by a differential equation.
Our goal in this section is to derive the differential
equation, and to understand the approximations that are 
made in deriving it.

The solution of equation (\ref{eq:randosc}) is
simply
\begin{equation}
\psi(t)=\psi(0)\cdot\exp(i A^t) \ ,
\label{eq:psiexact}
\end{equation}
where 
$A^t\equiv \int_0^t A(t')dt'$.
We expand as follows,
\begin{equation}
\psi(t)=\psi(0)\cdot(1+iA^t-(AA^t)^t+\ldots) \ ,
\label{eq:psi}
\end{equation}
where $(A A^t)^t\equiv \int_0^t A(t')\int_0^{t'}A(t'')dt''dt'$;
note that $(1/2)(A^t)^2=(AA^t)^t$ is an identity
for any $A(t)$. 
Equation (\ref{eq:psi}) may be thought of as an expansion in powers 
of $A$; it is only valid for short times after 
$t=0$.  

We evaluate $\left<\psi(t)\right>$ as follows.  Since $A(t)$ is unaffected by $\psi(0)$,
$A(t)$ and $\psi(0)$ are uncorrelated.\footnote{
In fact, $A(t>0)$ and $\psi(0)$ are slightly correlated: $\psi(0)$
is affected by $A(t_1<0)$, which is in turn correlated
with $A(t_2>0)$ as long as $t_2-t_1\lesssim \taucor$.  Nonetheless,
in the limit $\arms\taucor\ll 1$ that we are considering,
$\psi$ only changes by a small amount in the time $\taucor$. 
So the correlation between $A(0<t\lesssim \taucor)$ and $\psi(0)$
is unimportant for the evolution of $\psi(t)$.
} 
So
\begin{equation}
\left<\psi(t)\right>=\left<\psi(0)\right>\cdot(1+i\left<A^t\right>-\left<(AA^t)^t\right>+\left<\ldots\right>) \ .
\label{eq:meanpsi}
\end{equation}
Since $\left<A\right>=0$, we have $\left<A^t\right>=0$.  
Next, we consider $\left<(AA^t)^t\right>$.
The values of $A$ at two different
times ``statistically overlap'' 
only when the two times are separated by less than the
correlation time; i.e., $\left<A(t')A(t'+\Delta t)\right>$ is non-zero only if 
$\Delta t<\taucor$.  Thus $\left<AA^t\right>\simeq A_{\rm rms}^2\taucor$ when
$t\gtrsim\taucor$, and 
\begin{equation}
\left<(AA^t)^t\right>\simeq A_{\rm rms}^2\taucor t \ .
\label{eq:armstcorrt}
\footnote{
For this equation to be 
approximately valid, $A$ must decorrelate sufficiently
rapidly that
$\left<A(t')A(t'+\Delta t)\right>$ goes 
to zero faster than $1/\Delta t$ for 
large $\Delta t>\taucor$; otherwise, $\left<(AA^t)^t\right>$ rises faster than 
the first power of $t$, and our derivation is invalid.
\label{foot:corr}}
\end{equation}

Therefore
\begin{equation}
\left<\psi(t)\right>\simeq\left<\psi(0)\right>(1-A_{\rm rms}^2\taucor t) \  ,
\ \ \ \ \taucor\lesssim t \ll \taucor/(A_{\rm rms}\taucor)^2 \ .
\label{eq:meanpsi2}
\end{equation}
Equivalently,
\begin{equation}
{d\over {dt}}\left<\psi(t)\right>=-A_{\rm rms}^2\taucor\left<\psi(t)\right> \ ,
\label{eq:meanpsi3}
\ \ \ \ {\rm assuming}\ \ \arms\taucor\ll 1 \ .
\end{equation}
This equation is the main result
of this section.  
Van Kampen (1992) gives a
a more mathematically rigorous derivation of it than we do.
Equation (\ref{eq:meanpsi3}) can be understood as follows: since 
$\psi$ changes by $\Delta \psi\sim\arms\taucor \psi$ in the time $\taucor$,
then after $N$ steps, each $\taucor$ long, the change in $\psi$ is
$\sqrt{N} \Delta \psi$. 
Thus 
for order unity changes in $\psi$,
$(\psi/\Delta \psi)^2$ steps are required.
The resulting time, the ``cascade time,'' is 
\begin{equation}
\taucas=(\psi/\Delta \psi)^2\taucor\sim 1/(A_{\rm rms}^2\taucor) \ ,
\label{eq:taucas}
\end{equation}
as in equation (\ref{eq:meanpsi3}),
and $\taucor/\taucas\sim (\arms\taucor)^2\ll 1$.
Although $\left<\psi(t)\right>$ decays to zero on the timescale $\taucas$,
the energy $\vert\psi\vert^2$ remains constant; the ``cascade''
of $\left<\psi\right>$ is analogous to phase mixing.

We may now proceed to derive the kinetic equation in weak turbulence.
Before doing so, we consider the linear random oscillator in more detail.
If the reader is satisfied with the above derivation of equation (\ref{eq:meanpsi3}),
the following subsection may be skipped.

\subsubsection{The Validity of Perturbation Theory and Goldreich \& Sridhar (1997)}

Goldreich \& Sridhar (1997) incorrectly claim that perturbation theory fails in 
weak turbulence, i.e., in their ``intermediate'' turbulence.\footnote{
A note on terminology: the 
turbulence that we
and that Galtier et al. (2000) call ``weak,'' Goldreich \& Sridhar (1997)
call ``intermediate.'' They call it ``intermediate'' 
precisely because of their claim that perturbation theory
is invalid.}
We explain their claim and its resolution 
in the context of the linear random oscillator.
In the process, we clarify the validity of the perturbation expansion.

We consider the terms neglected in equation (\ref{eq:meanpsi2}).
For example, the fourth-order term is  
$\left<\psi^{(4)}\right>/\left<\psi(0)\right>=\left<\{A[A(AA^t)^t]^t\}^t\right>$. 
In this quadruple time integral, 
$A$ is evaluated at four different times.  Whenever two of
these times are separated by less than $\taucor$, the values
of $A$ at these two times ``statistically overlap,'' and hence 
can give a non-zero contribution to the total integral.  
If two of the times 
are separated by less than $\taucor$, and
the other two times are also separated by less than $\taucor$,
then the integrand does not vanish, even if
the first two times are separated from the second
two times by more than $\taucor$. So the quadruple time integral yields
approximately
$\left<\psi^{(4)}\right>/\left<\psi(0)\right>\sim (A_{\rm rms}^2\taucor t)^2=(t/\taucas)^2$. 

Since $\left<\psi^{(2)}\right>/\left<\psi(0)\right>=-A_{\rm rms}^2\taucor t=-t/\taucas$,
the contribution of the fourth-order term 
is as large as the second-order term
after the time $t=\taucas$.  Similarly, if we consider the
contribution to $\left<\psi^{(2n)}\right>/\left<\psi(0)\right>$ from correlating pairs
of $A$, the result is $\sim (t/\taucas)^n$; so all terms are
of comparable value when $t=\taucas$, and it seems that 
the lowest order term is inadequate.  This, in effect, is the claim
that Goldreich \& Sridhar (1997) make in the context of weak turbulence.

It is incorrect;
although equation (\ref{eq:meanpsi2}) for $\left<\psi(t)\right>$  
is only valid for $t\ll\taucas$,  
equation (\ref{eq:meanpsi3}) for $(d/dt)\left<\psi(t)\right>$ 
is approximately valid for all times, with corrections
of order powers of $\taucor/\taucas\ll 1$.
Since $\psi(t)$ undergoes small uncorrelated changes every
timestep of length $\taucor$, we expect on physical grounds that
the evolution of its statistical properties 
should be governed by a differential
equation that is invariant under time translations.\footnote{
It should be invariant on the timescale $\taucor$;
$\arms$ and $\taucor$ are effectively constants on this timescale.}
Equation (\ref{eq:meanpsi3}) is the 
only such equation whose small-time behaviour is given
by equation (\ref{eq:meanpsi2}). 
Its right-hand side may be interpreted as the lowest 
term in a perturbative expansion in $\taucor/\taucas$.
All of the contributions to $\left<\psi(t)\right>$ that 
are of order $(t/\taucas)^n$ must be derivable from equation
(\ref{eq:meanpsi3}). 
For example, if $A_{\rm rms}^2\taucor\equiv 1/\taucas$ 
is constant, then equation
(\ref{eq:meanpsi3}) has the solution 
$\left<\psi(t)\right>/\left<\psi(0)\right>=\exp(-t/\taucas)=
1-t/\taucas+(1/2)(t/\taucas)^2+\ldots$.  
Terms of order $(t/\taucas)^n$ in $\left<\psi\right>$ are 
generated from the lowest order term.

We can solve the one-dimensional oscillator exactly
when $A(t)$ is Gaussian, and 
thereby illustrate the validity of equation 
(\ref{eq:meanpsi3}).
If $A(t)$ 
is a zero-mean Gaussian random variable, then so is $A^t$, and
$\left<\exp(i A^t)\right>=\exp(-\left<(A^t)^2\right>/2)$.\footnote{
Proof: a zero-mean Gaussian $x$ 
has probability distribution 
$P(x)=(2\pi x_{\rm rms}^2)^{-1/2}\exp(-x^2/2x_{\rm rms}^2)$;
so $\left<\exp (ix)\right>=\int_{-\infty}^\infty P(x)\exp(ix)dx=
\exp(-x_{\rm rms}^2/2)$.}
It follows from equation (\ref{eq:psiexact}) that
\begin{equation}
\left<\psi(t)\right>=\left<\psi(0)\right>\cdot \exp(-\left<(AA^t)^t\right>) \ ,
\label{eq:psiexact2}
\end{equation}
assuming that $\left<\psi(0)\right>$ and $A(t)$ are uncorrelated.
Equivalently,
\begin{equation}
{d\over dt}\left<\psi\right>=-\left<AA^t\right>\left<\psi\right> \ ,
\label{eq:psiexact3}
\end{equation}
which agrees with equation (\ref{eq:meanpsi3}) when 
$\left<AA^t\right>=A_{\rm rms}^2\taucor$.

Although we are mainly concerned with the limit
$\arms\taucor\ll 1$, we conclude this subsection by briefly 
discussing the opposite
limit, $\arms\taucor\gg 1$.  
This limit sheds some light on strong turbulence, which
corresponds to the case $\arms\taucor \sim 1$. 
When  $\arms\taucor\gg 1$,
perturbation theory is inapplicable because
$\psi$ does not undergo small uncorrelated changes every $\taucor$.
Rather, its cascade time is $\sim 1/\arms$, which is much shorter
than the correlation time of $A$.  Therefore $A$ is nearly constant
in the time that $\left<\psi\right>$ cascades.  
The inapplicability of perturbation theory makes strong turbulence
difficult, if not impossible, to solve.
Nonetheless, we can solve the
one-dimensional oscillator in the ``strong'' limit.
Equations (\ref{eq:psiexact2}) 
and (\ref{eq:psiexact3}) are still valid, but
$\left<AA^t\right>\sim A_{\rm rms}^2 t$, so 
$\left<\psi(t)\right>\sim\left<\psi(0)\right>\exp(-A_{\rm rms}^2 t^2)$, and
equation (\ref{eq:psiexact3}) is not invariant under
time translations.
The time $t=0$ is special because of
our assumption that $\psi(0)$ and $A(t)$ are uncorrelated.
This assumption, while innocuous for the ``weak'' oscillator, is
crucial for the ``strong'' one. 

\subsection{Derivation of the Kinetic Equation}
\label{sec:derivke}

We derive the kinetic equation in Fourier space.
We Fourier transform equation (\ref{eq:eomrmhdappendix})
in $x$ and $y$ (but not $z$), denoting the 
transform of $\bwupp$ by $\bwupphk$:
\begin{equation}
\bwupphk\equiv \int d^2{\bld x_\perp} 
\bwupp
\exp(-i\bk\cdot{\bld x_\perp})  \ ,
\label{eq:fourier}
\end{equation}
where $\bk$ is purely transverse ($k_z\equiv 0$);
similarly, $\bwdownphk$ is the Fourier transform
of $\bwdownp$.

Since $\bwupphk$ is perpendicular to both $\bk$ and 
$\zhat$, it only represents a single degree of freedom.
Hence it is convenient to define a scalar 
potential $\psiupk$ by
$\bwupphk=i(\khat{\bld \times}\zhat)\psiupk$,
with $\khat\equiv \bk/k$.
Similarly, $\bwdownphk=i(\khat{\bld \times}\zhat)\psidownk$.
The Fourier transform of equation (\ref{eq:eomrmhdappendix})
is then
\begin{equation}
{\partial\over\partial t}\Big\vert_{\bk,\zup}
\psiupk(t)
=
\int d^2\bld{p} \ A_{\bk,\bp}(t) \psiupp(t) \ ,
\label{eq:fouriereqn}
\end{equation} 
where
\begin{eqnarray}
A_{\bk,\bp}(t)
&\equiv& a_{\bk,\bp}\psi^\downarrow_{\bk-\bp}(t)
\label{eq:fourier2a}
\\
&\equiv&
{1\over (2\pi)^2}
\zhat\cdot({\bld k\times \bld p})
\khat\cdot{\bld {\hat{p}}}
{\psi^\downarrow_{\bk-\bp}(t)\over
\vert\bk-{\bp}\vert}  \ .
\label{eq:fourier2}
\end{eqnarray}
We suppress the functional dependences 
of $\psi^\uparrow$ and $\psi^\downarrow$ on 
$\zup$ because
this equation is evaluated at fixed $\zup$;
in the following, we replace the partial time derivative
by a total derivative, with the understanding that $\zup$ 
is fixed.\footnote{
Since $\bwupp$ is real, $\bwupphk=(\bwupphmk)^*$,
where $*$ denotes the complex 
conjugate, and so $\psiupk=(\psiupmk)^*$; similarly,
$\psidownk=(\psidownmk)^*$. 
It follows that
$A_{\bk,\bp}=A_{-\bk,-\bp}^*$ and
$a_{\bk,\bp}=a_{-\bk,-\bp}^*$. 
The differential energy in up-waves within the 
$k$-space area 
$d^2\bk$ 
is proportional to 
$d^2\bk\vert\bwupphk\vert^2=d^2\bk\vert\psiupk\vert^2$.
Conservation of up-wave energy necessitates 
$A_{\bk,\bp}=-A^*_{\bp,\bk}$ and 
$a_{\bk,\bp}=-a_{\bp,\bk}^*$, as can be verified
from equations (\ref{eq:fourier2a}) and (\ref{eq:fourier2}). 
\label{foot:arelations}
}

We use angled brackets to denote an ensemble average,
in a plane with fixed $\zup$.  
We define the spectral energy densities
$e^\uparrow$ and $e^\downarrow$
as follows
\begin{eqnarray}
\left<\bwupphk(t)\bcdot\bwupphkp (t)\right>&=&
\left<\psi^\uparrow_\bk(t)\psi^\uparrow_\bkp(t)\right>=
\eupp(\bk,t)\delta(\bk+\bk') \label{eq:corrappendix} \\
\left<\bwdownphk(t)\bcdot\bwdownphkp (t)\right>&=&
\left<\psi^\downarrow_\bk(t)\psi^\downarrow_\bkp(t)\right>=
\edownp(\bk,t)\delta(\bk+\bk') \ ,
\label{eq:corrappendix2}
\end{eqnarray}
where $\eupp$ and $\edownp$ are real;
$\delta(\bk)$ is a two-dimensional Dirac delta-function
that follows from  homogeneity, i.e.,
from the assumption that $\left<\bwupp(\bxp)\cdot\bwupp(\bxp+\bdxp)\right>$ 
is independent of $\bxp$.

We shall derive the evolution equation for the
bilinear quantity $\left<\psiupk\psiupkp\right>$. From equation 
(\ref{eq:fouriereqn}),
\begin{equation}
{d\over dt} [\psiupk\psiupkp]
=\int d^2\bp A_{\bk,\bp} [\psiupkp\psiupp]
+(\bk\leftrightarrow\bkp) \ ,
\label{eq:eqnbilinear}
\end{equation}
where $(\bk\leftrightarrow\bkp)$ represents a second term that
is the same as the first, but
with $\bk$ and $\bkp$ interchanged.

From \S 
\ref{sec:prelim}, $\psi^\downarrow$ (and hence $A_{\bk,\bp}$)
can be viewed as evolving independently of $\psi^\uparrow$,
since the backreaction is negligible at fixed $\zup$. 
Therefore, equation (\ref{eq:eqnbilinear}) 
is nearly identical 
to that of the simple random oscillator
(eq. [\ref{eq:randosc}]); 
$[\psiupk\psiupkp]$ and $A_{\bk,\bp}$ in 
the above equation correspond to $\psi$ 
and $A$, respectively, in 
equation (\ref{eq:randosc}). 
Since $\psi^\downarrow$ is a random function
with zero mean
and correlation time $\taucor=\lpar/v_A$, so is $A_{\bk,\bp}$.
We solve equation (\ref{eq:eqnbilinear}) it in the same manner
as we solved the random oscillator,
using perturbation theory. 
Expanding in $A_{\bk,\bp}$, we find
to zeroth order:
\begin{equation}
[\psiupk\psiupkp]^{(0)}={\rm constant} \ .
\end{equation}
To first order:
\begin{equation}
[\psiupk\psiupkp]^{(1)}=
\int d^2\bp A^t_{\bk,\bp} [\psiupkp\psiupp]^{(0)}
+(\bk\leftrightarrow\bkp) \ ,
\end{equation} 
where $A^t_{\bk,\bp}\equiv\int_0^t A_{\bk,\bp}(t')dt'$.
To second order:
\begin{equation}
[\psiupk\psiupkp]^{(2)}=
\int d^2\bp d^2\bq \Big\{
(A_{\bk,\bp} A_{\bkp,\bq}^t)^t
[\psiupp\psiupq]^{(0)}
+
(A_{\bk,\bp} A_{\bp,\bq}^t)^t
[\psiupkp\psiupq]^{(0)}
\Big\}
+(\bk\leftrightarrow\bkp) \ .
\label{eq:psisecondorder}
\end{equation}
The sum of these last three equations is directly
analogous to equation 
(\ref{eq:psi}) for the simple oscillator.
Using the same reasoning here as we did for the
oscillator in deriving equation (\ref{eq:meanpsi3}),
and assuming that the correlation time ($\lpar/v_A$) is short,
we take the time derivative of the expected value
of equation (\ref{eq:psisecondorder});
we get, after
setting $\left<AA^t\right>=
(\lpar/v_A)\left<AA\right>$ (with the appropriate subscripts
on $A$), and after using 
equation (\ref{eq:corrappendix}) and integrating
out the delta functions on the right-hand side:
\begin{equation}
\delta(\bk+\bkp) {d\over dt}\eupp(\bk,t)=
{\lpar\over v_A}
\int d^2\bp
\Big\{
\left<A_{\bk,\bp} A_{\bkp,-\bp}\right>\eupp(\bp,t)
+\left<A_{\bk,\bp} A_{\bp,-\bkp}\right>\eupp(\bkp,t)
\Big\}
+(\bk\leftrightarrow\bkp) \ .
\label{eq:after}
\end{equation}
We re-express this equation by using
equations (\ref{eq:fourier2a}) and 
(\ref{eq:corrappendix2}) and the relations 
$a_{\bk,\bp}=a_{-\bk,-\bp}^*$~,
$a_{\bk,\bp}=-a_{\bp,\bk}^*$ 
(from footnote \ref{foot:arelations}), and
$\eupp(\bk,t)=\eupp(-\bk,t)$:
\begin{equation}
{d\over dt} \eupp(\bk,t) =2 {\lpar\over v_A}
\int d^2\bp \big(
\eupp(\bp,t)-\eupp(\bk,t)\big)
\vert a_{\bk,\bp}\vert^2\edownp(\bk-\bp,t) \ .
\end{equation}
Since each upgoing plane with fixed $\zup$ interacts with
many statistically independent downgoing wavepackets before
cascading, it is
reasonable to assume isotropy in planes transverse to $\zhat$;
after inserting equation (\ref{eq:fourier2}) for
$a_{\bk,\bp}$ and changing variables,
\begin{equation}
{d\over dt}
\eupp(k,t)=
{\lpar\over v_A}
k^2
\int_0^\infty dk_2 k_2^3
\Big[\eupp(k_2,t)-\eupp(k,t)\big]
\int_0^{2\pi} d\theta \sin^2\theta\cos^2\theta
\frac {\edownp(k_1,t)}{k_1^2} \ ,\label{eq:kineticequationappendix}
\end{equation}
where
\begin{equation}
k_1\equiv (k^2+k_2^2-2kk_2\cos\theta)^{1/2} \ .
\end{equation}
We have redefined $\lpar$ to absorb the factor of $4\pi^3$, i.e., 
$\lpar/(4\pi^3)\rightarrow\lpar$.
Thus far, we have only considered a single plane with fixed $\zup$.
If we assume that the turbulence is homogeneous in $z$, the $z$-average
of the above equation is trivial since there is no explicit
dependence on $z$.  We can simply re-interpret angular
brackets to denote $z$-averages in addition to an ensemble average.

We emphasize that 
our derivation of the kinetic 
equation (eq. [\ref{eq:kineticequationappendix}])
is only valid when the correlation time, $\lpar/v_A$, is 
much smaller than the cascade time, $(v_A/\lpar)(k^4\eupp)^{-1}$.
Otherwise, equation (\ref{eq:after}) 
does not follow from equation (\ref{eq:psisecondorder}).

\subsection{Steady State Fluxes}
\label{sec:steadystatefluxes}

In steady state, we set the right-hand side of 
equation (\ref{eq:kineticequationappendix}) to zero.
We substitute power law solutions, $\eupp(k)\propto k^{-(3+\alphaupup)}$,
$\edownp(k)\propto k^{-(3+\alphadown)}$, in which case the
right-hand side becomes
\begin{eqnarray}
\Big[
\eupp(k_0)\edownp(k_0)k_0^{6+\alphaupup+\alphadown}
{\lpar\over v_A}
\Big]
k^2
\int_0^k d\kup \kup^3
\Big(\kup^{-(3+\alphaupup)}-k^{-(3+\alphaupup)}\Big)
\Big(1-(\kup/k)^{\alphaupup+\alphadown}\Big) \cdot \nonumber\\
\int_0^{2\pi} d\theta \sin^2\theta\cos^2\theta
\Big(
k^2+\kup^2-2k\kup\cos\theta
\Big)^{-(5+\alphadown)/2} \ .
\label{eq:fluxexpression}
\end{eqnarray}
Note that the
square-bracketed term is 
independent of $k_0$.
The above expression follows 
after breaking the $\kup$ integral into two pieces: one
from 0 to $k$, and the second from $k$ to $\infty$.  
Then, the change of
variables 
$\kup\rightarrow k^2/\kup$
is made in the second piece
(a ``Zakharov transformation''),
so that its limits are now from 0 to $k$.  Finally, this second
piece is combined with the first.  

In steady state, equation (\ref{eq:fluxexpression}) must vanish, so
\begin{equation}
\alphadown=-\alphaupup \ . \label{eq:alpha}
\end{equation}
Since
$\wuppl\sim (k^2\eupp)^{1/2}\propto k^{-(1+\alphaupup)/2}$, and similarly
for $\wdownpl$, equation (\ref{eq:alpha}) is equivalent to
$\wuppl\wdownpl\propto\lambda$ (eq. 
[\ref{eq:intscal2}]).  
The vanishing of $\pt\edownp$ in steady state yields
the same relation as equation (\ref{eq:alpha}), and so does not
give new information.

The flux associated with $\eupp$ is given by 
integrating the right-hand side of equation 
(\ref{eq:kineticequationappendix}) over $d^2\bk/(2\pi)$ from 
a particular $k$ to infinity.
The steady state flux is thus given by integrating 
equation (\ref{eq:fluxexpression}), with $\alphadown=-\alphaupup$.
The result of the flux integration is
\begin{equation}
\epsilon^\uparrow=f(\alphaupup)\Big[\eupp(k_0)\edownp(k_0)k_0^6\frac{\lpar}{v_A}\Big] \ ,
\label{eq:flux1}
\end{equation}
where $f(\alphaupup)$ is a dimensionless function
of $\alphaupup$:  
\begin{equation}
f(\alphaupup)=\int_0^1
dx x^3\ln x \big( x^{-(3+\alphaupup)}-1\big)
\int_0^{2\pi} d\theta \sin^2\theta\cos^2\theta
\big(
1+x^2-2x\cos\theta
\Big)^{-(5-\alphaupup)/2} \ ,
\label{eq:fappendix}
\end{equation}
A technical
note: although equation 
(\ref{eq:fluxexpression})
vanishes in steady state, its integral over $d^2\bk$ gives a factor
of $\alphaupup+\alphadown$ in the denominator, which also vanishes
in steady state.  This $0/0$ ambiguity can be resolved by 
considering the limit as
$\alphaupup+\alphadown\rightarrow 0$ (L'H\^{o}pital's rule).

Equation (\ref{eq:flux1}) and the corresponding equation for $\epsilon^\downarrow$,
are equivalent to those in the heuristic
discussion 
(eqs. [\ref{eq:epsf1}] and [\ref{eq:epsf2}]).

Galtier et al. (2000) plot the function $f(\alphaupup)$ after numerically
integrating the steady-state flux integral.\footnote{More precisely,
their figure 2 is proportional to
$[f(\alphaupup)f(-\alphaupup)]^{-1/4}$, and
their figure 3 shows $f(\alphaupup)/f(-\alphaupup)$.}
They show 
that $-1<\alphaupup<1$, 
that the steeper spectrum always carries
more flux (i.e., $f(\alphaupup)/f(-\alphaupup)$ is a
monotonically increasing function of $\alphaupup$), 
and that in the limit that
$\alphaupup\rightarrow 1$, 
$\epsilon^\uparrow/\epsilon^\downarrow\rightarrow \infty$.
(Clearly, this also implies that in the limit $\alphaupup\rightarrow -1$,
$\epsilon^\uparrow/\epsilon^\downarrow\rightarrow 0$.)  
In \S \ref{sec:locality}, we explain the physical reason
why $-1<\alphaupup<1$.  When this condition is violated, the
cascade becomes nonlocal.  Galtier et al. (2000)
find infinite fluxes when these inequalities are saturated
because they consider an infinitely extended spectrum, which
leads to unphysical results when the cascade is nonlocal.

As discussed in \S \ref{sec:fixedfluxes}, 
in steady state
we are primarily concerned with the
case that the up- and down-going fluxes are comparable, 
so $\vert\alphaupup\vert \ll 1$. 
In this limit, we linearize $f$ about $\alphaupup=0$,
yielding
approximately 
\begin{equation}
f(\alphaupup)\simeq f(0)\cdot (1+0.5\alphaupup) \ \ , \ \vert\alphaupup\vert\ll 1 \ ,
\label{eq:pt5}
\end{equation}
after numerical integrations of equation (\ref{eq:fappendix}).
This result is used in \S \ref{sec:fixedfluxes}
to calculate the energy spectra given the fluxes $\epsilon^\uparrow$
and $\epsilon^\downarrow$.
Although the value $f(0)$ is much less important than the dependence
of $f$ on $\alphaupup$, we use it when discussing the results of
our numerical simulations; it is given by
\begin{equation}
f(0)\simeq 1.87 \ .
\label{eq:f0}
\end{equation}
 
\subsection{The Zero-Frequency Mode}
\label{sec:compare}

Weak MHD turbulence is often described as being based
on three-wave resonances,
in which one of the three waves has zero frequency
(e.g.,
Shebalin, Matthaeus, \& Montgomery 1983,
Galtier et al. 2000, 2002).  
However, the interpretation of a zero-frequency wave
is unclear.  After all, it should require an infinite amount
of time for three interacting waves to ``realize'' that one
of them has zero frequency. 
In this subsection, we 
clarify the role of the zero-frequency mode.  
In the process, we shall compare our derivation of the kinetic
equation with that of Galtier et al. (2002).
Note that 
in our derivation in \S \ref{sec:derivke}
we avoided discussing Fourier wavemodes, since we did not Fourier-transform
in $t$ or in $z$.

The evolution of the up-waves at fixed $\zup\equiv z-v_A t$ is
given in equation (\ref{eq:eqnbilinear}).
However, for clarity, in this subsection 
we shall consider
the random oscillator instead (eq. [\ref{eq:randosc}]):
\begin{equation}
d\psi/dt=iA\psi \ .
\end{equation}
It is a simple matter to extend our discussion to weak MHD 
turbulence by
replacing $\psi$ with  
$\psiupk\psiupkp$, replacing $iA$ with $A_{\bk,\bp}$~, and
summing over the appropriate indices.

In \S \ref{sec:toy} we derived the 
equation for $\left<\psi\right>$, assuming that
the cascade time of $\left<\psi\right>$ is longer than
the correlation time of $A$: 
\begin{equation}
d\left<\psi\right>/dt=-\left<AA^t\right>\left<\psi\right> \ ,
\label{eq:psiav}
\end{equation}
where $\left<AA^t\right>\equiv \int_{t-T}^t\left<A(t) 
A(t')\right>dt'
\sim A_{\rm rms}^2\taucor$; the value of $T$ 
is unimportant as long as $T>\taucor$.

We can see how the zero-frequency mode enters by
re-writing $\left<AA^t\right>$ in terms of 
$\hat{A}_\omega$, the
Fourier transform of $A$: 
\begin{equation}
\left<AA^t\right>
=\int_{-\infty}^\infty {d\omega'\over 2\pi}
\int_{-\infty}^\infty {d\omega\over 2\pi}
\left<\hat{A}_{\omega'}\hat{A}_\omega\right>
e^{i\omega't}\int_{t-T}^t e^{i\omega t'} dt' \ .
\label{eq:aat}
\end{equation}
To compare this result with that of Galtier et al. (2002),
we shall use similar notation.
We denote 
the power spectrum of 
$\hat{A}_\omega$ by  
$q_\omega$:
\begin{equation}
\left<\hat{A}_{\omega'}\hat{A}_\omega\right>\equiv q_\omega \delta(\omega'+\omega) \ ;
\end{equation} 
the Dirac-delta function results from the time-invariance of the statistical properties
of $A$.
We also define
\begin{equation}
\Delta(\omega)\equiv {1\over i\omega}e^{i\omega t}(1-e^{i\omega T}) \ .
\end{equation}
With these two definitions, equation (\ref{eq:aat}) is
\begin{equation}
\left<AA^t\right>=
\int_{-\infty}^\infty {d\omega\over 4\pi^2}
q_\omega \Big[e^{-i\omega t}\Delta(\omega)\Big] \ .
\end{equation}
If this equation is inserted into equation (\ref{eq:psiav}),
the resulting equation is equivalent to equation (7) in Galtier et al. (2002).
As these authors argue, when $T\rightarrow\infty$, 
$\exp(-i\omega t)\Delta(\omega)\rightarrow \pi\delta(\omega)$,
so
\begin{equation}
\left<AA^t\right>=
\int_{-\infty}^\infty {d\omega\over 4\pi}q_\omega \delta(\omega) \ ,
\end{equation}
which corresponds to  
equation (8) in Galtier et al. (2002).
From this expression, we can see why the zero-frequency mode enters:
it is simply a consequence of the term $\left<AA^t\right>$.  
It does not require an infinite time for $\left<\psi\right>$ to interact with 
$q_{\omega=0}$; rather, the only quantity that enters into $q_{\omega=0}$
is the value of $A$ within the time $\taucor$ of $t$. 
We can see this more clearly by explicitly writing the expression for $q_\omega$. 
Of course, the power spectrum $q_\omega$ is simply the Fourier-transform
of the correlation function of $A$:
\begin{equation}
q_\omega
=4\pi\int_{-\infty}^0 e^{-i\omega\tau}\left<A(t)A(t+\tau)\right> d\tau \ . 
\label{eq:qomega}
\end{equation}
As long as $\omega\lesssim 1/\taucor$, we have $q_\omega\simeq 
4\pi\left<AA^t\right>$.

One of our reasons for discussing the Fourier-space picture
in detail is that there have been a large
number of confusing remarks about it in the literature.  
For example, 
Galtier et al. (2000, 2002)
claim that,
since the kinetic equation is apparently not applicable to
the zero-frequency mode, 
there might be a ``condensation'' of zero-frequency modes.  
Since the zero-frequency mode is so important, this condensation
might have dramatic implications for the cascade.
However, as
long as correlation times are
shorter than cascade times, the kinetic equation gives a complete
description of the turbulence, and is applicable to the
zero-frequency mode as well, which is
simply given by $q_{\omega=0}=4\pi\left<AA^t\right>$.
For this not to be true, large-time correlations would have to 
build up in $A$ (see eq. [\ref{eq:qomega}]). 
Since $A$ corresponds to $\psi^\downarrow$ in weak MHD turbulence,
large-time correlations at fixed $\zup$ correspond to large-distance 
correlations along the $z$-direction.
But it is impossible for two downgoing wavepackets, 
initially 
uncorrelated, to become correlated.
This is because the equation for their evolution is linear: recall that in weak
turbulence the backreaction term is negligible.

\end{document}